\documentclass[aps,amsfonts,pra,singlecolumn,superscriptaddress,showpacs]{revtex4}
\usepackage{epsfig,amsmath,amssymb,bm,epsf,graphics}
\def\bra#1{\langle#1\vert}
\def\ket#1{\vert#1\rangle}
\def\ketbra#1{\vert#1\rangle\langle#1\vert}

\def\ipr#1#2{\langle#1\vert#2\rangle}

\def\Longarrow{\protect\@lra}
\def\@lra{\relbar\joinrel\relbar\joinrel\relbar\joinrel%
          \relbar\joinrel\rightarrow}

\def\coe#1{{({\rm co}E_{#1})}}
\begin{document}
\title{Relative entropy of entanglement
       for certain multipartite mixed states}
 \author{Tzu-Chieh Wei}
 \affiliation{
    Institute for Quantum Computing and
    Department of Physics and Astronomy,
    University of Waterloo,
    200 University Avenue West,
    Ontario, Canada N2L 3G1}
\date{\today}
\begin{abstract}
We prove conjectures on the relative entropy of entanglement (REE) for two
families of multipartite  qubit states. Thus, analytic expressions of REE for
these families of states can be given. The first family of states are composed
of mixture of some permutation-invariant multi-qubit states. The results
generalized to multi-qudit states are also shown to hold. The second family of
states contain D\"ur's bound entangled states. Along the way, we have
discussed the relation of REE to two other measures: robustness of
entanglement and geometric measure of entanglement, slightly extending
previous results.
\end{abstract}
 \maketitle
\section{Introduction}
 Entanglement, the characteristic trait of quantum mechanics according to Schr\"odinger,
 has been identified as a resource central to
many of quantum information processing tasks~\cite{NielsenChuang00}. There
have been tremendous progress along the characterization and the
quantification of entanglement using various methods in both bipatite and
multipartite settings~\cite{Horodecki4}. Entanglement has also been studied in
many-body systems and connection to quantum phase transitions has also been
explored~\cite{AmicoFazioOsterlohVedral08}. The notion of entanglement has
also found its application in density matrix of renormalization groups (DMRG)
and has enabled much recent progress in numerical techniques of DMRG in low
dimensions and dynamics~\cite{Schollwock05}.

Despite these advancements, the characterization and the quantification of
entanglement are far from complete. For example, even for multipartite pure
states, it has been a long standing question whether there exists a finite
minimal reversible entanglement generating set
(MREGS)~\cite{BennettPopescuRohrlichSmolinThapliyal}. The existence of MREGS
would enable the generalization of the entanglement of
distillation~\cite{BennettBernsteinPopescuSchumacher96} and
formation~\cite{BennettDiVincenzoSmolinWootters96,Wootters98} to multipartite
systems~\cite{PlenioVedral01} and would also provide a better characterization
of multipartite entanglement. Given that the standard measures of
entanglement, such as the entanglement of
formation~\cite{BennettDiVincenzoSmolinWootters96,Wootters98} and
distillation~\cite{BennettDiVincenzoSmolinWootters96,Wootters98} have not be
properly generalized to multipartite settings, the study of quantifying
multipartite entanglement via other measures is indispensable.

The relative entropy of entanglement (REE), introduced by Vedral et
al.~\cite{VedralPlenioRippinKnight97,VedralPlenio98}, provides an alternative
measure of entanglement. In bipartite settings, it is shown to be a lower
bound for the entanglement of formation and an upper bound for the
entanglement of distillation. Its regularized version has recently been shown
to possess connection to the second law of
thermodynamics~\cite{BrandaoPlenio07}. Moreover, REE applies straightforwardly
to multipartite settings, and thus its study in the multipartite settings is
important.
 The calculation of REE~\cite{VedralPlenioRippinKnight97,VedralPlenio98} by analytic means,
however, is still a challenging task, even for {\it pure\/} states, and is
often relied on numerical computation. Despite the progress made in two-qubit
systems by Ishizaka~\cite{Ishizaka03}, an analytic formula for REE is still
elusive.

In this paper, we shall focus on the study of REE, complemented by two other
measures. These other measures, the geometric measure of entanglement (based
on the geometry of Hilbert space~\cite{Shimony95,BarnumLinden01,WeiGoldbart03}
and also based on the Grover
search~\cite{MiyakeWadati01,BihamNielsenOsborne02}, hence, also known as the
Groverian entanglement) and the generalized robustness of
entanglement~\cite{VidalTarrach99}, besides providing different perspectives
of entanglement on their own, can also be used to provide connections to REE,
or more precisely, lower and upper bounds, respectively. The main results of
this paper are the establishment of the proof for conjectures for analytic
expressions of REE for several families of multipartite mixed states, thus
providing nontrivial examples where REE is obtained analytically.

 The structure of the present paper is as follows. In Sec.~\ref{sec:measures} we
review the three entanglement measures considered in the paper: the relative
entropy of entanglement, the geometric measure of entanglement, and the
robutsness of entanglement. We explore connections among the three, in both
pure- and mixed-state settings in Sec.~\ref{sec:connectionpure}  and
Sec.~\ref{sec:connectionmixed}, respectively. In Sec.~\ref{sec:mixture} we
prove the conjecture for multi-qubits~\cite{WeiEricssonGoldbartMunro04} and we
also generalize the result to the multi-qudit setting. With these analytic
results, we discuss some applications in Sec.~\ref{sec:applications}. In
Sec.~\ref{sec:Dur} we prove the analytic formula of REE for another family of
multi-qubit mixed states~\cite{WeiAltepeterGoldbartMunro04}, which include
D\"ur's bound entangled states~\cite{Dur01}. In Sec.~\ref{sec:summary} we give
some concluding remarks.

\section{Entanglement measures}
\noindent%
\label{sec:measures}%
In this section we briefly review the three measures considered in the present
paper: the relative entropy of entanglement, the general (global) robustness
of entanglement and the geometric measure of entanglement.
\subsection{Relative entropy of entanglement}
\noindent
The relative entropy $S(\rho||\sigma)$ between two states
$\rho$ and $\sigma$ is defined via
\begin{equation}
S(\rho||\sigma)\equiv {\rm Tr}\left(\rho\log_2\rho-\rho\log_2{\sigma}\right),
\end{equation}
which is evidently not symmetric under exchange of $\rho$ and $\sigma$, and is
non-negative, i.e., $S(\rho||\sigma)\ge 0$. Note that the $\log$ function is
base-2 throughout this paper. The relative entropy of entanglement (RE) for a
mixed state $\rho$ is defined to be the minimal relative entropy of $\rho$
over the set of separable mixed
states~\cite{VedralPlenioRippinKnight97,VedralPlenio98}:
\begin{equation}
\label{eqn:ER}
E_{\rm R}(\rho)\equiv \min_{\sigma\in {\cal D}}S(\rho||\sigma)=\min_{\sigma\in {\cal D}}{\rm Tr}\left(\rho\log_2\rho-\rho\log_2\sigma\right),
\end{equation}
where ${\cal D}$ denotes the set of all separable states.

The regularized relative entropy of entanglement is defined as
\begin{equation}
E_{\rm R}^\infty(\rho)\equiv \lim_{n\rightarrow \infty}\frac{1}{n}E_{\rm
R}(\rho^{\otimes n}).
\end{equation}
It is shown that in the bipartite settings the regularized relative entropy of
entanglement plays an analogous role to the entropy in
thermodynamics~\cite{BrandaoPlenio07}. However, the calculation of the
regularized relative entropy of entanglement is, in general, much more
difficult than for the non-regularized case.

In general, the task of finding the REE for arbitrary states $\rho$ involves a
minimization over all separable states, and this renders the computation of
the REE very difficult.  For bipartite pure states, the REE is equal to the
entanglement of formation and of distillation. But, despite recent
progress~\cite{Ishizaka03}, for mixed states---even in the simplest setting of
two qubits---no analog of Wootters' formula~\cite{Wootters98} for the
entanglement of formation has been found. Things are even worse in
multipartite settings.  Even for pure states, there has not been a systematic
method for calculating their relative entropy of entanglement. It is thus
worthwhile seeking cases in which one can explicitly obtain an expression for
the REE.

An alternative definition of RE is to replace the set of separable states
${\cal D}$ by the set of postive partial transpose (PPT) states ${\cal
D}^{ppt}$:
\begin{equation}
{\cal D}^{ppt} =\{\sigma|| \,\sigma^\dagger=\sigma, \ \sigma\ge 0, \ {\rm Tr}
(\sigma)=1, \ \sigma^{PT}\ge 0\},
 \end{equation}
 where $PT$ denotes partial transpose with respect to any bi-partition of
 parties. The REE thus defined, as well as its regularized version, gives a tighter bound
on distillable entanglement. There has been important progress in calculating
the RE (and its regularized version) with respect to PPT states for certain
bipartite mixed states; see Refs.~\cite{AudenaertEtAl} for more detailed
discussions. For multipartite settings one could also use this definition, and
define the set of states to optimize over to be the set of states that are PPT
with respect to certain or  all bipartite partitionings. However, we shall use
the first definition, i.e., optimization over the set of completely separable
states, throughout the discussion of the present paper.

\subsection{General robustness of entanglement}
The general robustness of entanglement~\cite{VidalTarrach99} is a measure of
how sensitive the entanglement is to mixture of states. It is defined as
\begin{eqnarray}
R(\rho)\equiv\min t
\end{eqnarray}
such that there exists a state $\Delta$ so as to render the following state
separable:
\begin{equation}
\frac{1}{1+t}(\rho+ t\Delta).
\end{equation}
The logarithmic robustness of $\rho$ is defined as
\begin{equation}
LR(\rho)\equiv \log_2 \Big(1+ R(\rho)\Big).
\end{equation}

\subsection{Geometric measure of entanglement}
\noindent We continue by briefly reviewing the formulation of the geometric
measure in both pure-state and mixed-state settings. Let us start with a
multipartite system comprising $n$ parts, each of which can have a distinct
Hilbert space. Consider a general $n$-partite pure state (expanded in the
local bases $\{|e_{p_i}^{(i)}\}$):
\begin{equation}
|\psi\rangle=\sum_{p_1\cdots p_n}\chi_{p_1p_2\cdots p_n}
|e_{p_1}^{(1)}e_{p_2}^{(2)}\cdots e_{p_n}^{(n)}\rangle.
\end{equation}
We can compare this state to the set of  general separable pure state,
\begin{equation}
\ket{\phi}\equiv\mathop{\otimes}_{i=1}^n|\phi^{(i)}\rangle
=\mathop{\otimes}_{i=1}^{n}
\Big(\sum_{p_i}c_{p_i}^{(i)}\,|e_{p_i}^{(i)}\rangle\Big),
\end{equation}
and define
 the maximal overlap of $\ket{\psi}$ with the closest product states as
 follows,
\begin{equation}
\Lambda_{\max}({\psi})=\max_{\phi}|\ipr{\phi}{\psi}|,
\end{equation}
where $\ket{\phi}$ is an arbitrary separable pure state defined above. In
Ref.~\cite{WeiGoldbart03}, the particular form $E_{\sin^2}\equiv
1-\Lambda^2_{\max}({\psi})=\sin^2\theta_{\min}$ was defined to be the
geometric measure of entanglement (GME) for any pure state $|\psi\rangle$.
Here, we shall be concerned with the related quantity
$E_{\log}(\psi)\equiv-2\log_2\Lambda_{\max}({\psi})$, which we shall show to
be related to the two other measures.

We remark that an alternative perspective from the Grover search also leads to
the geometric measure~\cite{MiyakeWadati01,BihamNielsenOsborne02}, and it is
also known as the Groverian entanglement. Furthermore, a hierarchy of
entanglement can be obtained if one consider the separable states to be
product states among an appropriate partition of particles into
$k$-parties~\cite{BarnumLinden01,WeiGoldbart03}; see recent works by Shimoni
and Biham~\cite{ShimoniBiham07} and by Blasone et
al.~\cite{BlasoneDellAnnoDeSienaIlluminati07}. But throughout this paper we
shall focus on the completely product states, and the separable mixed states
will refer to $n$-separable states.

Given the definition of entanglement for pure states just formulated,
the extension to mixed states $\rho$ can be built upon pure states
via the {\it convex hull\/} construction (indicated by ``co''),
as was done for the entanglement of formation; see Ref.~\cite{Wootters98}.
 The essence is a minimization
over all decompositions $\rho=\sum_i p_i\,|\psi_i\rangle\langle\psi_i|$
into pure states:
\begin{eqnarray}
\label{eqn:Emixed}
E(\rho)
\equiv
\coe{\rm pure}(\rho)
\equiv
{\min_{\{p_i,\psi_i\}}}
\sum\nolimits_i p_i \,
E_{\rm pure}(|\psi_i\rangle).
\end{eqnarray}
This convex hull construction ensures that the measure gives zero for
separable states; however, in general it also complicates the task of
determining mixed-state entanglement. The specific form that we are concerned
in this paper is the following
\begin{equation}
\label{eqn:Elog}
 E_{\log}(\rho) \equiv {\min_{\{p_i,\psi_i\}}}
\sum\nolimits_i p_i \big[-2\log\Lambda_{\max}(\psi_i)\big].
\end{equation}

We remark that an alternative way to define the mixed-state entanglement
measure via purification have been developed by Shapira et
al.~\cite{ShapiraShimoniBiham07}, continuing along the idea of maximizing the
Groverian search outcome. But in the present paper, we shall focus mainly on
the definition by Eq.~(\ref{eqn:Elog}).

\smallskip
\noindent {\it Illustrative examples\/}: We examine some pure-state examples,
whose mixture will be considered later in the paper. First, one can classify
permutation-invariant pure states, as follows:
\begin{equation}
\label{eqn:Snk} |S(n,k)\rangle\equiv \frac{1}{\sqrt{C^n_k}} {\cal S
}|\underbrace{0\cdots0}_{k}\underbrace{1\cdots1}_{n-k}\rangle=
\frac{\sqrt{C^n_k}}{n!}\sum_i
\Pi_i|\underbrace{0\cdots0}_{k}\underbrace{1\cdots1}_{n-k}\rangle,
\end{equation}
where ${\cal S}$ represents symmetrization under permutations, $\Pi_i$ denotes
arbitrary permutation of $n$ objects, and $C^n_k\equiv n!/{k!(n-k)!}$.
Intuitively, as the amplitudes are all positive, one can assume that the
closest separable (equivalently, Hartree) state is of the form (which is
rigorously proved in Ref.~\cite{HayashiMarkhamMuraoOwariVirmani08})
\begin{equation}
\label{eqn:separable}
|\phi\rangle=\big(\sqrt{p}\,|0\rangle+\sqrt{1-p}\,|1\rangle\big)^{\otimes n},
\end{equation}
for which the maximal overlap (w.r.t.~$p$) gives the entanglement
eigenvalue for $|{\rm S}(n,k)\rangle$:
\begin{eqnarray}
\label{eqn:Lambda}
\Lambda_{\max}(n,k)=
\sqrt{\frac{n!}{k!(n\!-\!k)!}}
\left(\frac{k}{n}\right)^{\frac{k}{2}}
{\left(\frac{n-k}{n}\right)}^{\frac{n\!-\!k}{2}}.
\end{eqnarray}
More generally, for $n$ parties each of which being a $d$-level system, the
state
\begin{equation}
\label{eqn:Snkk} \ket{S(n;\vec{k})}\equiv \sqrt{\frac{k_1!k_2!\cdots
k_d!}{n!}}\, {\cal S}\,
|\underbrace{1\ldots1}_{k_1}\,\underbrace{2\dots2}_{k_2}\ldots
 \underbrace{d\ldots d\,}_{k_d}\,\rangle
\end{equation}
has the entanglement eigenvalue
\begin{equation}
\Lambda_{\max}(n;\vec{k})=\sqrt{\frac{n!}{\prod_i (k_i!)}} \,\prod_{i=1;k_i\ne
0}^{d}\left(\frac{k_i}{n}\right)^{\frac{k_i}{2}}.
\end{equation}

Although the above states were discussed in terms of the
GME~\cite{WeiGoldbart03}, we shall, in the following section, show the rather
surprising fact that the LR and REE of these example states, are given by the
corresponding expression: $-2 \log_2\Lambda_{\max}$.

One of the main results of the present paper concerns the two following
families of mixture of symmetric states:
\begin{eqnarray}
\label{eqn:snk}\rho(\{\vec{p}\})&=&\sum_{k=0}^n p_k\ketbra{S(n,k)}, \\
\label{eqn:snveck}\rho(\{p_{\vec{k}}\})&=&\sum_{\vec{k}}
p_{\vec{k}}\ketbra{S(n,\vec{k})}.
\end{eqnarray}

\section{Connection between the three measures: pure states}
\label{sec:connectionpure} In bipartite systems, due to the existence of
Schmidt decompositions, the relative entropy of entanglement of a pure state
is simply the von Neumann entropy of its reduced density matrix. However, for
multipartite systems there is, in general, no such decomposition, and how to
calculate the relative entropy of entanglement for an arbitrary pure state
remains an open question. We now connect the three measures by inequalities.

\subsection{
Relative entropy of entanglement and geometric measure of entanglement}
 Let us begin with the following inequality:

\smallskip
\noindent {\bf Inequality 1\/}~\cite{WeiEricssonGoldbartMunro04,VidalEtAl}:
For any pure state $\ket{\psi}$ with entanglement eigenvalue
$\Lambda_{\max}(\psi)$ the quantity $-2\log_2\Lambda_{\max}(\psi)$ is a lower
bound on the relative entropy of entanglement of $\ket{\psi}$, i.e.,
\begin{equation}
\label{eqn:2log}
E_{\rm R}(\ketbra{\psi})\ge -2\log_2\Lambda_{\max}(\psi).
\end{equation}

\smallskip\noindent
{\bf Proof\/}: From the definition~(\ref{eqn:ER}) of the relative entropy of
entanglement we have, for a pure state $\ket{\psi}$,
\begin{equation}
E_{\rm R}(\ketbra{\psi})=
 \min_{\sigma\in {\cal D}}-\bra{\psi}\log_2\sigma\ket{\psi}=
-\max_{\sigma\in {\cal D}} \bra{\psi}\log_2\sigma\ket{\psi}.
\end{equation}
Using the concavity of the log function, we have
\begin{equation}
\label{eqn:inequalityLog}
\bra{\psi}\log_2\sigma\ket{\psi}\le \log_2 (\bra{\psi}\sigma\ket{\psi})
\end{equation}
and, furthermore,
\begin{equation}
\label{eqn:inequalityMax} \max_{\sigma\in{\cal D}}
\bra{\psi}\log_2\sigma\ket{\psi}\le \max_{\sigma\in{\cal D}} \log_2
(\bra{\psi}\sigma\ket{\psi}).
\end{equation}
 We then conclude that
\begin{equation}
E_{\rm R}(\ketbra{\psi})\ge
-\max_{\sigma\in{\cal D}}
\log_2 (\bra{\psi}\sigma\ket{\psi}).
\end{equation}
As any $\sigma\in{\cal D}$ can be expanded as
$\sigma=\sum_i p_i\ketbra{\phi_i}$,
where ${\ket{\phi_i}}$'s are separable
pure states, one has
\begin{equation}
\bra{\psi}\sigma\ket{\psi}=
\sum_i p_i |\ipr{\phi_i}{\psi}|^2 \le
\Lambda^{2}_{\max}({\psi}),
\end{equation}
and hence we arrive at the sought result
\begin{equation}
E_{\rm R}(\ketbra{\psi})\ge
-2\log_2\Lambda_{\max}({\psi}).
\label{eq:theorem}
\end{equation}
{\hfill \ensuremath{\Box}}

\subsection{Geometric measure and generalized robustness}
 Cavalcanti has obtained the result that robustness of entanglement can be shown to provide
 an upper for the geometric measure~\cite{Cavalcanti06}. The insight he provided is the connection of both measures to
 entanglement witness. We shall repeat his result for pure states here.
 The connection of the generalized robustness to entanglement witness is the result of
 Brand\~ao~\cite{Brandao05}:
 \begin{equation}
 \label{eqn:Brandao}
 R(\rho)=\max \{0, -\min_{W} {\rm Tr}(W\rho)\},
 \end{equation}
 where $M\le \openone$ and ${\rm Tr}(\rho_{SEP})\ge0$. Here we shall be concerned with pure states, and
 hence
 \begin{equation}
 R(\ketbra{\psi})=\max \{0, -\min_{W} \bra{\psi}W\ket{\psi})\}.
 \end{equation}
The connection of the geometric measure of entanglement to entanglement
witness is as follows. The operator
\begin{equation}
W=(\lambda^2\openone-\ketbra{\psi})/\lambda^2
\end{equation}
satisfies the condition ${\rm Tr}(W\rho_{SEP})\ge 0$  as long as $\lambda^2\ge
\Lambda^2_{\max}(\psi)$~\cite{WeiGoldbart03}. Using this and take
$W^*=(\Lambda_{\max}(\psi)^2\openone-\ketbra{\psi})/\Lambda^2_{\max}(\psi)$,
it will provide a lower bound on $R(\ketbra{\psi})$, and thus
\begin{equation}
R(\ketbra{\psi})\ge  \frac{1-\Lambda^2_{\max}{\psi}}{\Lambda^2_{\max}(\psi)}.
\end{equation}
Equivalently, we have the following \\
{\bf Inequality 2\/}~\cite{Cavalcanti06}:
\begin{equation}
LR(\ketbra{\psi})\ge  -\log_2{\Lambda^2_{\max}(\psi)}.
\end{equation}
It turns out that a stronger lower bound on the robustness can be obtained
using the relative entropy of entanglement, first proved by Hayashi et
al.\cite{HayashiMarkhamMuraoOwariVirmani06}, which we now describe.
\subsection{ Generalized robustness and relative entropy of entanglement}
We repeat here the proof that for any
pure state $\ket{\psi}$ (with $\rho_\psi\equiv \ketbra{\psi}$), \\
{\bf Inequality 3\/}~\cite{HayashiMarkhamMuraoOwariVirmani06}:
\begin{equation}
E_R(\ketbra{\psi})\le LR(\ketbra{\psi}).
\end{equation}
{\bf Proof\/}. Suppose $R(\rho_\psi)=t_m$. There exists a state $\Delta$ such
that the following state is separable
\begin{equation}
\sigma_S=\frac{1}{1+t_m}\Big(\ketbra{\psi}+t_m\Delta\Big).
\end{equation}
Thus we have
\begin{eqnarray}
E_R(\ketbra{\psi})\le
S\Big(\ketbra{\psi}\Big\vert\Big\vert\sigma_S\Big)&=&-\bra{\psi}\log\frac{\ketbra{\psi}+t_m\Delta}{1+t_m}\ket{\psi}\nonumber\\
&\le& -\bra{\psi}\log\frac{\ketbra{\psi}}{1+t_m}\ket{\psi}
=-\log\frac{1}{1+t_m}=LR(\ketbra{\psi}).
\end{eqnarray}
{\hfill \ensuremath{\Box}}

 Inequalities~1 to 3 summarize to \\
 \noindent {\bf Inequality 3'\/}:\\
\begin{equation}
LR(\ketbra{\psi})\ge E_{\rm R}(\ketbra{\psi})\ge
-2\log_2\Lambda_{\max}({\psi}). \label{eq:theorem2}
\end{equation}
The degree of difficulty of calculating the corresponding measures becomes
higher from the r.h.s. to the l.h.s. Moreover, upper bounds on $LR$ and
$E_{\rm R}$ are easy to obtain, as they are defined via minimization. If one
can construct an upper bound on these measures such that it matches
$-2\log_2\Lambda_{\max}({\psi})$, the inequalities above become equalities.
This is how the analytic expression of left two measures have been obtained
for symmetric and antisymmetric states as well as some graph
states~\cite{WeiEricssonGoldbartMunro04,HayashiMarkhamMuraoOwariVirmani06,HayashiMarkhamMuraoOwariVirmani08,MarkhamMiyakeVirmani07}.
A very thorough analysis using group theory on whether the inequalities become
inequality has been provided by Hayashi et
al.~\cite{HayashiMarkhamMuraoOwariVirmani08}.
\subsection{Illustrative examples}
 We now examine illustrative states in the light of the
above discussions, thus obtaining the expression of the three different
entanglement measures. We begin with the permutation-invariant states
$\ket{S(n,k)}$ of Eq.~(\ref{eqn:Snk}), for which $\Lambda_{\max}$ was given in
Eq.~(\ref{eqn:Lambda}). The above theorem guarantees that $E_{\rm
R}\big(\ket{S(n,k)}\big)\ge -2\log_2 \Lambda_{\max}(n,k)$. To find an {\it
upper\/} bound we construct a separable mixed state
\begin{subequations}
\begin{eqnarray}
\sigma^*
&\equiv&
\int \frac{d\phi}{2\pi}\ketbra{\xi(\phi)},
\\
\ket{\xi(\phi)}
&\equiv&
\left(\sqrt{p}\ket{0}+
e^{i\phi}\sqrt{1-p}\ket{1}\right)^{\otimes n},
\end{eqnarray}
\end{subequations}
with $p$ chosen to maximize
$||\ipr{\xi}{S(n,k)}||=\sqrt{C_k^n \,p^k(1-p)^{n-k}}$, which gives $p=k/n$.
Direct evaluation then gives
\begin{equation}
\label{eqn:sigmastar2} \sigma^*=\sum_{j=0}^n C^n_j
p^{j}(1-p)^{(n\!-\!j)}\ketbra{S(n,j)}=\Lambda^2_{\max}(n,k)\ketbra{S(n,k)}+\tau^\perp,
\end{equation}
and $S(\rho||\sigma)=-2\log_2\Lambda_{\max}(n,k)$, where
$\rho=\ketbra{S(n,k)}$, $\Lambda_{\max}(n,k)$ is given in
Eq.~(\ref{eqn:Lambda}), and we have used $\tau^\perp$ to denote the those
terms with $(j\ne k$). The upper and lower bounds on $E_{\rm R}$ coincide, and
hence we have that
\begin{equation}
\label{eqn:rhoSnk}
E_{\rm R}\big(\ket{S(n,k)}\big)=-2\log_2 \Lambda_{\max}(n,k).
\end{equation}

Moreover, from the perspective of the generalized robustness, we immediately
have an upper bound $t_m$ on the robustness,
\begin{equation}
\frac{1}{1+t_m}= \Lambda^2_{\max}(n,k).\end{equation} Hence, we obtain an
upper bound on $LR$,
\begin{equation}
 LR(n,k)\le \log_2 (1+t_m) =-2\log_2 \Lambda_{\max}(n,k).
\end{equation}
This then gives equality to all three measures for the symmetric state
$\ket{S(n,k)}$. The same consideration also holds for $\ket{S(n,\vec{k})}$,
namely,
\begin{equation}
\label{eqn:equalityqudit} LR(n,\vec{k})=E_{\rm R}(n,\vec{k})=-2\log_2
\Lambda_{\max}(n,\vec{k}).
\end{equation}

\section{Connection among the three measures: mixed states}
\label{sec:connectionmixed}  Cavalcanti and Hayashi et al. have provided
inequalities relating these three measures for general mixed states. In their
inequalities, the geometric measure is generalized to
\begin{equation}
G(\rho)\equiv- \log_2 \{\max_{\sigma\in{\cal D}}{\rm Tr}(\rho\sigma)\}.
\end{equation}
We shall see below that inequalities in terms of $E_{\log}$ can also be
derived and shall provide an alternative inequality.

Let us first review the inequality shown by  Cavalcanti~\cite{Cavalcanti06}:\\
{\bf Inequality 4\/}~\cite{Cavalcanti06}:
\begin{equation}
LR(\rho)\ge G(\rho).
\end{equation} Recall that we take ${\cal D}$
to be the set of separable states of the form
\begin{equation}
\sum_i p_i \ketbra{\phi_i^A}\otimes\ketbra{\phi_i^B}\otimes\dots,
\end{equation}
and hence $G(\rho)$ can be simplified to be
\begin{equation}
G(\rho)=- \log_2 \{\max_{\phi\in{\rm Prod}}\bra{\phi}\rho\ket{\phi}\},
\end{equation}
where Prod denotes completely product states. \\
{\bf Proof\/}: Again, we use the relation~(\ref{eqn:Brandao}) for the
robustness and the entanglement witness obtained by
Brand\~ao~\cite{Brandao05}. For any state $\rho$, construct a witness
\begin{equation}
W_\lambda=\openone - \frac{1}{\lambda^2}\rho,
\end{equation}
where $\lambda^2\ge \lambda^2_{\max}\equiv \max_{\sigma\in{\cal D}} {\rm
Tr}(\rho\,\sigma)$, in order for the condition ${\rm Tr}(W_\lambda\sigma)\ge
0$ to be satisfied. Plug in the above expression for witness into
Eq.~(\ref{eqn:Brandao}), we obtain
\begin{equation}
R(\rho)\ge -{\rm Tr}(W_{\lambda_{\max}})=-1+\frac{1}{\lambda_{\max}^2}.
\end{equation}
This is equivalent to
\begin{equation}
LR(\rho)\ge -\log_2{\lambda_{\max}^2}=G(\rho).
\end{equation}{\hfill \ensuremath{\Box}}

The inequality derived by
 Hayashi
et al. is as follows, \\
{\bf Inequality 5\/}~\cite{HayashiMarkhamMuraoOwariVirmani06}:
\begin{equation}
\log_2 r(\rho)\ge E_R(\rho)+S(\rho)\ge G(\rho),
\end{equation}
where, instead of the robustness of $\rho$ itself, the robustness of the
support of $\rho$ is considered:
\begin{equation}
r(\rho)\equiv |P_\rho| [1+R(P_\rho/|P_\rho|)],
\end{equation}
where $P_\rho$ is the support of $\rho$, namely
\begin{equation}
P_\rho=\sum_{\lambda_i\ne 0} \ketbra{\lambda_i},
\end{equation}
where $\rho$ has the spectral decomposition
$\rho=\sum_i\lambda_i\ketbra{\lambda_i}$, and $|P_\rho|\equiv {\rm
Tr}(P_\rho)$.

We now slightly extend the first part of the inequality. By the definition of
robustness, there exist a state $\Delta$ such that the state
\begin{equation}
\omega= \frac{{\rho}+t\Delta}{1+t}
\end{equation}
is a separable state, where $t=R(\rho)$ for convenience. This gives
\begin{eqnarray}
E_R({\rho})+S({\rho})&=&\min_{\sigma\in{\cal D}} - {\rm
Tr}(\rho\log\sigma) \nonumber \\
&\le& - {\rm Tr}(\rho\log\omega)\nonumber\\
&=& - {\rm Tr}\left(\rho\log\frac{\rho+t\Delta}{1+t}\right)\nonumber\\
&\le&- {\rm Tr}\left(\rho\log\frac{\rho}{1+t}\right)\nonumber\\
 &=&S(\rho)+LR(\rho).
\end{eqnarray}
This gives
\begin{equation}
LR(\rho)\ge E_R(\rho),
\end{equation}
where the l.h.s. is given by the logarithmic robustness of $\rho$, rather than
that of its support $P_\rho/ |P_\rho|$. The second inequality of Hayashi et
al. is elementary to prove:
\begin{equation}
E_R({\rho})+S({\rho})=-\max_{\sigma\in{\cal D}}  {\rm Tr}(\rho\log\sigma) \ge
- \log\big\{\max_{\sigma\in{\cal D}}{\rm Tr}(\rho\,\sigma)\big\},
\end{equation}
where we have used $ {\rm Tr}(\rho\log\sigma) \le \log{\rm Tr}(\rho\,\sigma)$.

We can also  provide an alternative of the second-part of Inequality 5:
\begin{equation}
 E_R(\rho)\ge E_{\log}(\rho)-S(\rho).
\end{equation}
\noindent {\bf Proof\/}: \\
Suppose $\rho=\sum_i p_i\ketbra{\psi_i}$ is the
optimal decomposition for $E_{\log}(\rho)$, namely,
\begin{equation}
E_{\log}(\rho)= -\sum_i p_i \log_2 \Lambda_{\max}^2(\psi_i).
\end{equation}
Using the definition of REE, we have
\begin{equation}
E_R({\rho})+S({\rho})=-\max_{\sigma\in{\cal D}}  {\rm Tr}(\rho\log\sigma) \ge
-\max_{\sigma_i\in{\cal D}} \sum_i p_i {\rm Tr}(\ketbra{\psi_i}\log\sigma_i).
\end{equation}
Using \begin{equation} \max_{\sigma_i\in{\cal D}}{\rm
Tr}(\ketbra{\psi_i}\log\sigma_i)=\log\Lambda_{\max}^2(\psi_i),
\end{equation}
we have
\begin{equation}
E_R({\rho})+S({\rho}) \ge- \sum_i p_i\log
\Lambda_{\max}^2(\psi_i)=E_{\log}(\rho) .
\end{equation} {\hfill \ensuremath{\Box}}\\
Summing up we have proved \\
\noindent {\bf Inequality 6\/}:
\begin{equation}
LR(\rho)\ge E_R({\rho})\ge E_{\log}(\rho)-S({\rho}) .
\end{equation}

Let us compare Inequality 6 to Inequality 5, especially the first parts.
Viewed as an upper bound on $E_{\rm R}$, the Inequality 5 can be rewritten as
\begin{equation}
\log_2r(\rho)-S(\rho)\ge E_{\rm R}(\rho).
\end{equation}
It can happen that the r.h.s. is zero whereas the l.h.s. is still nonzero, as
exemplified for the two-qubit Werner state,
\begin{equation}
\rho_W(\gamma)=\gamma\ketbra{\psi^-}+\frac{1-\gamma}{4}\openone_{4\times 4},
\end{equation}
where $\ket{\psi^-}$ is the two-qubit singlet state. The state is entangled
for  $\gamma>1/3$, and unentangled for $\gamma\le 1/3$. For $0\le \gamma <1$,
the support of $\rho_W$ is $P_W=\openone_{4\times 4}$, and the corresponding
state is a completely mixed state, hence, possessing no entanglement. This
leads to $r(\rho_W)=4$, and
\begin{equation}
S(\rho_W(\gamma))=-\frac{1+3\gamma}{4}\log_2\frac{1+3\gamma}{4}-\frac{3(1-\gamma)}{4}\log_2\frac{1-\gamma}{4}.
\end{equation}
At $\gamma=1/3$, $E_{\rm R}(\rho_W)=0$, but
$\log_2r(\rho)-S(\rho)=1-\log_2(3)/2\approx 0.208$. Although $LR(\rho)$ is
generally not easy to calculate, it becomes zero when $\rho$ becomes
separable. For the second-part of Inequality 6, we remark that for the states
in Eqs.~(\ref{eqn:snk}), (\ref{eqn:snveck}) and (\ref{eqn:DurW}) the lower
bound can be tightened to
\begin{equation}
E_R(\rho)\ge E_{\log}(\rho),
\end{equation}
which will be shown later. Therefore, Inequality 6 slightly extends previous
results by Cavalcanti~\cite{Cavalcanti06} and Hayashi et
al.~\cite{HayashiMarkhamMuraoOwariVirmani06}.
\section{Relative entropy of entanglement for mixture of symmetric states}
\label{sec:mixture}
\subsection{Multi-qubits}
 In Ref.~\cite{WeiGoldbart03} the procedure was given to find the geometric
measure of entanglement, $E_{\sin^2}$, for the mixed state comprising
symmetric states:
\begin{equation}
\label{eqn:mixSnk}
\rho(\{p\})=\sum_k p_k\,\ketbra{S(n,k)}.
\end{equation}
Here, we focus instead on the quantity $E_{\log}$, but the basic procedure is
the same.  The key point is to find the entanglement eigenvalue
$\Lambda_{n}(\{q\})$ for the pure state
\begin{equation}
\sum_k \sqrt{q_k}\,\ket{S(n,k)},
\end{equation}
thus arriving at the quantity
\begin{equation}
\label{eqn:epsilon}
{\cal E}(\{q\})\equiv -2 \log_2 \Lambda_{n}(\{q\}).
\end{equation}
Then the quantity $E_{\log}$ for the mixed state~(\ref{eqn:mixSnk}) is
actually the convex hull of the expression~(\ref{eqn:epsilon}):
\begin{equation}
E_{\log_2}\left(\rho(\{p\})\right)={\rm co}\,{\cal E}(\{p\}).
\end{equation}

In Ref.~\cite{WeiEricssonGoldbartMunro04}, an attempt to calculate the REE for
the states~(\ref{eqn:mixSnk}) was made and a conjecture for the REE was made.
In this section, we review the construction of the conjecture and prove it to
be correct. Let us consider the state formed by mixing the separable pure
states $\ket{\xi(\theta,\phi)}$:
%\begin{subequations}
\begin{eqnarray}
\label{eqn:sigmatheta}
\sigma(\theta)=\int \frac{d\phi}{2\pi} \ketbra{\xi(\theta,\phi)}=\sum_{k=0}^n C^n_k\cos^{2k}\theta\sin^{2(n\!-\!k)}\theta\ketbra{S(n,k)},
\end{eqnarray}
%\end{subequations}
where
\begin{equation}
\label{eqn:xi}
\ket{\xi(\theta,\phi)}\equiv
\left(\cos\theta \ket{0}+ e^{i\phi}\sin\theta\ket{1}\right)^{\otimes n}.
\end{equation}
Allowing $\theta$ to vary, we then minimize the relative entropy between
$\rho(\{p\})$ and $\sigma(\theta)$,
\begin{equation}
S\left(\rho(\{p\})\vert\vert\sigma(\theta)\right)=\sum_k p_k
\log\frac{p_k}{C^n_k\cos^{2k}\theta\sin^{2(n\!-\!k)}\theta},
\end{equation}
with respect to $\theta$.  We arrive at the stationarity condition
\begin{equation}
\label{eqn:theta}
\tan^2\theta\equiv \frac{\sum_k p_k\,(n-k)}{\sum p_k \,k}.
\end{equation}
Due to the convexity of the relative entropy, namely,
\begin{equation}
S\left(\sum_i q_i \rho_i\Vert\sum_i q_i \sigma_i\right)
\le \sum_i q_i S(\rho_i||\sigma_i),
\end{equation}
we can further tighten the expression of the relative entropy by taking its
convex hull. The convexification process also results in the corresponding
separable state, i.e., the knowledge of the coefficients $r_k$'s in $\sigma^*$
\begin{equation}
\label{eqn:sigmastar}
 \sigma^*=\sum_k r_k\ketbra{S(n,k)}.
\end{equation}
The arrived upper bound on REE for the mixed state $\rho(\{p\})$ was
conjectured (shall be proved below) in Ref.~\cite{WeiEricssonGoldbartMunro04}
to be the exact REE:\\
{\bf Theorem 1\/}:
\begin{equation}
\label{eqn:conjecture} E_{\rm R}\left(\rho(\{p\})\right)= {\rm co}\,F(\{p\}),
\end{equation}
where
%\begin{subequations}
\begin{eqnarray}
\label{eqn:F}
F(\{p\})\equiv\sum_k p_k
\log_2 \frac{p_k}{C_k^n \cos^{2k}\theta \sin^{2(n-k)}\theta}=\sum_k p_k \log_2 \frac{p_k\, n^n}{C_k^n \alpha^{k} (n-\alpha)^{n-k}},
\end{eqnarray}
%\end{subequations}
where the angle $\theta$ satisfies Eq.~(\ref{eqn:theta}),
$C_k^n\equiv n!/\big(k!(n-k)!\big)$, and $\alpha\equiv \sum_k p_k\, k$.

\subsection{Proof of the theorem}
\subsubsection{Symmetry considerations}
We shall discuss the symmetries possessed by the states $\rho(\{p\})$, and
these shall reduce the set of separable states that we need to consider. We
begin by noting that the states $\rho(\{p\})$ are invariant under the
projection
\begin{equation}
\label{eqn:Projection} {\bf \rm P}_1:\rho\rightarrow \int\frac{d\phi}{2\pi}\,
U(\phi)^{\otimes n}\rho\, U(\phi)^{\dagger\otimes n}
\end{equation}
with $U(\phi)\big\{\ket{0},\ket{1}\big\}\to
\big\{\ket{0},{\rm e}^{-i\phi}\ket{1}\big\}$.
Vollbrecht and Werner~\cite{VollbrechtWerner01}
have shown that in order to find the closest separable mixed state
for a state that is invariant under projections such as ${\bf \rm P}$,
it is only necessary to search within the separable states that are
also invariant under the projection.

We can further reduce the set of separable states to be searched by invoking
another symmetry property possessed by $\rho(\{p\})$: these states are also,
by construction, invariant under permutations of all parties.  Let us denote
by $\Pi_i$ one of the permutations of parties, and by $\Pi_i(\rho)$ the state
obtained from $\rho$ by permuting the parties under $\Pi_i$.  We now show that
the set of separable states to be searched can be reduced to the separable
states that are invariant under the permutations. To see this, suppose that
$\rho$ is a mixed state in the family~(\ref{eqn:mixSnk}), and that $\sigma^*$
is one of the closest separable states to $\rho$, i.e.,
\begin{equation}
E_{\rm R}(\rho)\equiv\min_{\sigma\in {\cal D}} S(\rho||\sigma)=S(\rho||\sigma^*).
\label{eqn:extreme}
\end{equation}
As $\rho$ is invariant under all $\Pi_i$, we have
\begin{equation}
E_{\rm R}(\rho)=\frac{1}{N_\Pi}\sum_i S\left(\rho\big\Vert\Pi_i(\sigma^*)\right),
\end{equation}
where $N_\Pi=n!$ is the number of permutations. By using the convexity of the
relative entropy we have
\begin{equation}
E_{\rm R}(\rho)\ge S\left(\rho\big\Vert\big[\sum_i \Pi_i(\sigma^*)/N_\Pi\big]\right).
\end{equation}
However, because of the extremal property, Eq.~(\ref{eqn:extreme}), the
inequality must be saturated, as the left-hand side is already minimal. This
shows that the state under the projection
\begin{equation}
{\rm P}_2:\sigma^{*}\rightarrow \frac{1}{N_\Pi}\sum_i \Pi_i(\sigma^*)
\end{equation}
is also a closest separable mixed state to $\rho$, and is manifestly invariant
under all permutations (noting that ${\rm P}_2$ preserves separability).
Thus, we only need to search within this restricted family of separable
states, namely separable states invariant under ${\rm P}_1$ and ${\rm P}_2$.

In fact, one can use the group representation theory (see, e.g.
Ref.~\cite{KeylWerner01,HayashiMarkhamMuraoOwariVirmani08}) to characterize
all states that invariant under  ${\rm P}_1$ and ${\rm P}_2$. It turns out
that any such state (invariant under both ${\rm P}_1$ and ${\rm P}_2$) can be
written as mixture of all $\ket{S(n,k)}$ and those other basis states
belonging to other irreducible representations for the group $S_n$. However,
we have not been able to prove our theorem using only symmetry argument, for
it is not trivial to characterize all separable states of this form. But we
can prove it by an algebraic approach already developed by Vedral and
Plenio~\cite{VedralPlenio98}.

It is not difficult to see that the set ${\cal D}_S$ of all separable mixed states
that are diagonal in the basis of $\{\ket{S(n,k)}\}$ (basis states for the totally symmetric subspace ${\cal D}_S$) can be constructed from
a convex mixture of separable states in Eq.~(\ref{eqn:sigmatheta}).
That is, for any $\sigma_s\in {\cal D}_S$ we have a decomposition
\begin{equation}
\label{eqn:sigmas}
\sigma_s=\sum_i t_i \,\sigma(\theta_i),
\end{equation}
where $t_i\ge 0$, $\sum_i t_i=1$, and $\sigma(\theta_i)$ is of the form~(\ref{eqn:sigmatheta}).
This is because the separability of the states~(\ref{eqn:mixSnk}) implies
that there exists a decomposition into pure states such that
each pure state is a separable state. Furthermore,
 because $\{\ket{S(n,k)}\}$ are eigenstates of $\rho(\{p\})$,
the most general
form of the pure state in its decomposition is
\begin{equation}
\sum_k\sqrt{q_k}\,e^{i\phi_k}\ket{S(n,k)}.
\end{equation}
This pure state is separable if and only if it is of the form~(\ref{eqn:xi}),
up to an overall irrelevant phase. As $\rho(\{p\})$ is invariant under the
projection ${\rm P}_1$~(\ref{eqn:Projection}), a pure state in
Eq.~(\ref{eqn:xi}) will be projected to the mixed state in
Eq.~(\ref{eqn:sigmatheta}) under ${\rm P}_1$. Thus, every separable state that
is diagonal in $\{\ket{S(n,k)}\}$ basis can be expressed in the
form~(\ref{eqn:sigmas}).

Hence, our construction of $\sigma^*$ (via any necessary convexification)
ensures  that it achieves at least the minimum (of the relative entropy) when
the separable mixed states are restricted to ${\cal D}_S$. However, in order
to prove the conjecture, one would still need to show that the expression is
also the minimum when the restirction to ${\cal D}_S$ is relaxed to the set of
separable states invariant under ${\rm P}_1$ and ${\rm P}_2$, unless we can
employ further argument, symmetry or not, to reduce to ${\cal D}_S$ the set of
separable states that we need to consider.

\subsubsection{The proof}
The symmetry considerations presented in the previous section only reduce the
set of separable states to a smaller one possessing symmetry. In particular,
 any state that is invariant
under (1) any permutation of the $n$ parties and (2) projection under average
of local phase as in Eq.~(\ref{eqn:Projection}) can be written as a mixture of
the basis states in all the irreducible representations of $U(2)$. However,
the symmetry alone has not led us  to the proof of our theorem. Here, we
provide a more direct algebraic proof.

 Since we have shown that $\sigma^*$ gives minimum of
$S(\rho||\sigma)$ when the separable states are restricted to being diagonal
in $\ket{S(n,k)}$'s, we now show that it is indeed a local minimum when we
lift that restriction. According to the discussions by Vedral and Plenio~\cite{VedralPlenio98}, this means that we need to show that adding any
separable state $\sigma_s$: $\sigma(x)=(1-x)\sigma^* + x \sigma_s$, the
quantity $f$,
\begin{equation}
f(x,\sigma_s)\equiv S(\rho|| (1-x)\sigma^*+ x \sigma_s),
\end{equation}
has a local minimum at $x=0$. This is equivalent to show that
\begin{equation}
\frac{\partial f}{\partial x}(0,\sigma_s)\ge 0.
\end{equation}
It is straightforward to calculate the l.h.s. of the above expression, as was done in Ref.~\cite{VedralPlenio98}, which
gives
\begin{equation}
\frac{\partial f}{\partial x}(0,\sigma_s)=1-\int_0^\infty dt\, {\rm
Tr}\big[\rho (\sigma^*+t)^{-1}\sigma_s (\sigma^*+t)^{-1}\big].
\end{equation}
Taking $\sigma^*$ as in Eq.~(\ref{eqn:sigmastar}) and $\rho$ in
Eq.~(\ref{eqn:mixSnk}), we have
\begin{equation}
\frac{\partial f}{\partial x}(0,\sigma_s)=1-\sum_{k=0}^n
\frac{p_k}{r_k}\bra{S(n,k)}\sigma_s\ket{S(n,k)}.
\end{equation}
As $\sigma^*$ is constructed to be the minimum in the restricted (i.e., the
totally symmetric) subspace, this means that for $\ket{\Phi_s}=
(\cos\theta\ket{0}+\sin\theta\ket{1})^{\otimes n}$ in this subspace, we
automatically have by construction
\begin{equation}
\frac{\partial f}{\partial x}(0,\ketbra{\Phi_s})=1-\sum_k\frac{p_k}{r_k} C^n_k
\cos^{2k}\theta\sin^{2(n-k)}\theta\ge 0.
\end{equation}

As $\partial f(0,\sigma)/\partial x$ is linear in $\sigma$, namely,
\begin{equation}
\frac{\partial f}{\partial x}(0,\sum_i p_i\sigma_i)=\sum_i p_i \frac{\partial f}{\partial x}(0,\sigma_i),
\end{equation}
in order to show that ${\partial f}(0,\sigma)/{\partial x}\ge0$ holds for arbitrary separable state $\sigma$, it is sufficient to show that it holds for arbitrary pure separable state $\sigma=\ketbra{\Phi}$, which is what we are about to do.
To be more explicit, we shall show that for any arbitrary separable pure state
\begin{equation}
\ket{\Phi}=\otimes_{j=1}^n(\sqrt{q_j}\ket{0}+ \sqrt{1-q_j}
e^{i\phi_j}\ket{1}),
\end{equation}
we also have
\begin{equation}
\frac{\partial f}{\partial x}(0,\ketbra{\Phi})=1-\sum_{k=0}^n
\frac{p_k}{r_k}\ipr{S(n,k)}{\Phi}\ipr{\Phi}{S(n,k)}\ge 0.
\end{equation}
Note that $\ket{\Phi}$ is the most general separable multi-qubit pure state
(up to an irrelevant global phase).

\begin{figure}%[t]
\centerline{\psfig{figure=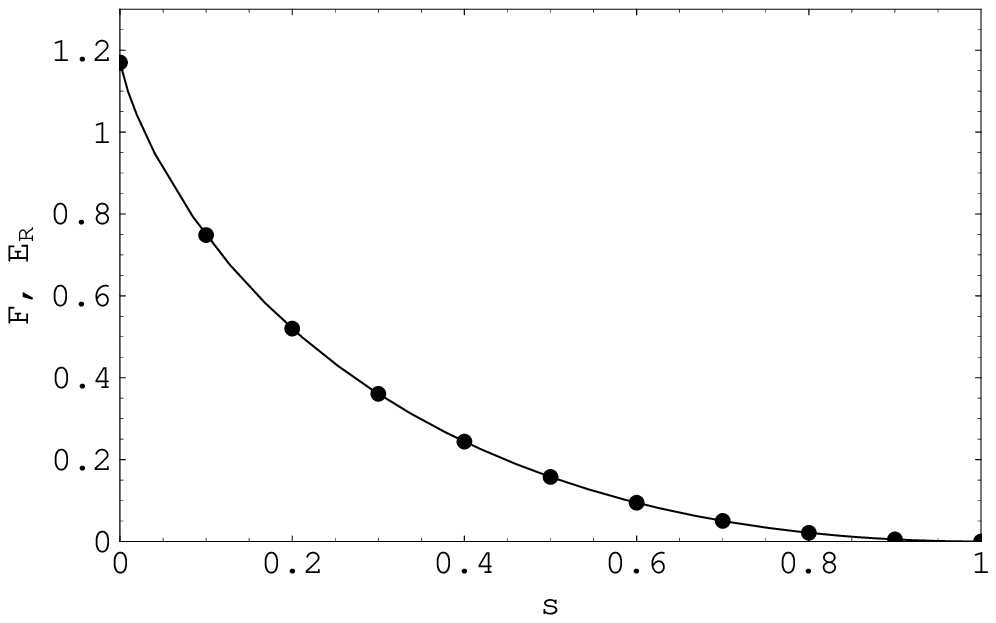,width=7cm,height=4cm,angle=0}}
\vspace{0.2cm}
\centerline{\psfig{figure=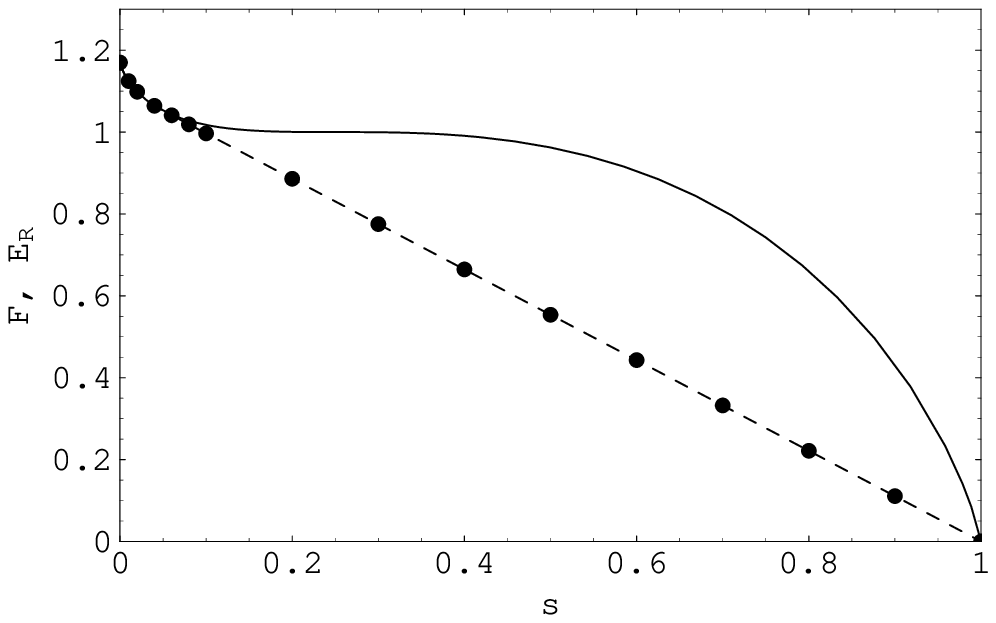,width=7cm,height=4cm,angle=0}}
\vspace{0.2cm}
\centerline{\psfig{figure=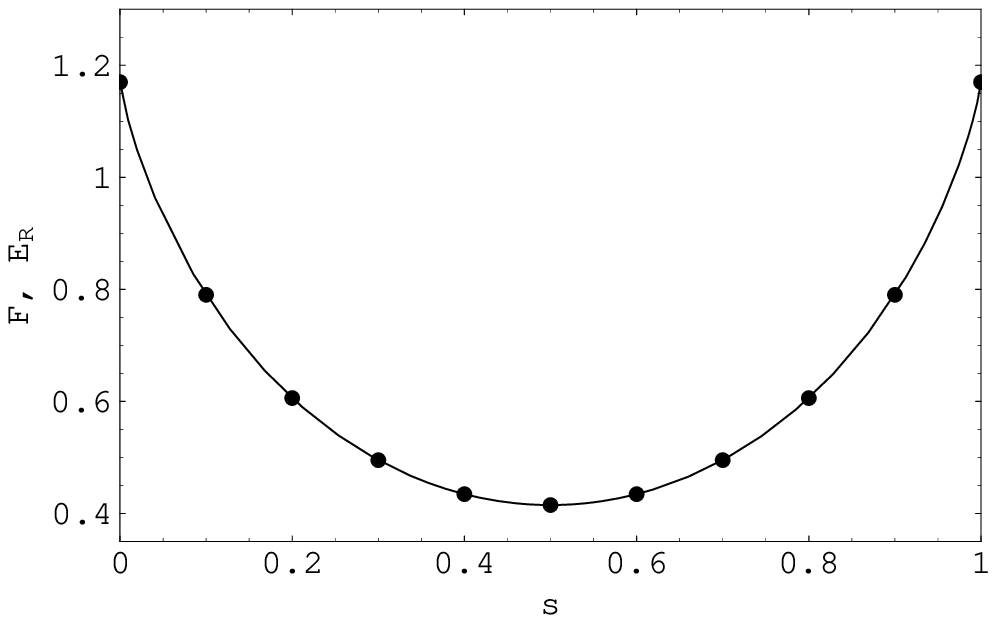,width=7cm,height=4cm,angle=0}}
%\vspace{- 0.2cm}
\caption{Comparison of $F$ (solid curve), $E_{\rm R}={\rm co}\,F$ (partly
dashed line, when convexification is necessary) and the numerical value of
$E_{\rm R}$ (dots) for the states $\rho_{3;0,1}(s)$, $\rho_{3;0,2}(s)$, and
$\rho_{3;1,2}(s)$ (from top to bottom). } \label{fig:Er3}
%\vspace{-0.5cm}
\end{figure}

The key point is then to evaluate
\begin{equation}
\Phi_k\equiv \ipr{S(n,k)}{\Phi}.
\end{equation}
Note that
\begin{equation}
\sqrt{C^n_k}\ket{S(n,k)}=\frac{1}{k!(n-k)!}\sum_{\Pi_i \in S_n}
\ket{\Pi_i(\underbrace{0,..,0}_k,\underbrace{1,..1)}_{n-k}}.
\end{equation}
This gives
\begin{equation}
\Phi_k\equiv
\ipr{S(n,k)}{\Phi}=\frac{\sqrt{C^n_k}}{n!}\sum_{\Pi_i}\sqrt{q_{\Pi_i(1)}}\dots\sqrt{q_{\Pi_i(k)}}
\sqrt{1-q_{\Pi_i(k+1)}}e^{i\phi_{\Pi_i(k+1)}}\dots\sqrt{1-q_{\Pi_i(n)}}e^{i\phi_{\Pi_i(n)}}.
\end{equation}
Thus,
\begin{equation}
|\Phi_k|\le\frac{\sqrt{C^n_k}}{n!}\sum_{\Pi_i}\sqrt{q_{\Pi_i(1)}}\dots\sqrt{q_{\Pi_i(k)}}
\sqrt{1-q_{\Pi_i(k+1)}}\dots\sqrt{1-q_{\Pi_i(n)}},
\end{equation}
and
\begin{equation}
|\Phi_k|^2\le\frac{C^n_k}{(n!)^2}\big[\sum_{\Pi_i}q_{\Pi_i(1)}\dots
q_{\Pi_i(k)}\big]\big[\sum_{\Pi_i}
(1-q_{\Pi_i(k+1)})\dots(1-q_{\Pi_i(n)})\big],
\end{equation}
where we have used the Cauchy-Schwarz inequality. Using the Maclaurin
inequality~\cite{HardyLittlewoodPolya}:
\begin{equation}
\frac{1}{n!}\sum_{\Pi_i} x_{\Pi_i(1)} \dots x_{\Pi_i(k)} \le
\left(\frac{\sum_{i=1}^n x_i}{n}\right)^k,
\end{equation}
we have that
\begin{equation}
|\Phi_k|^2\le C_k^n (\bar{q})^{k} (1-\bar{q})^{n-k}= C^n_k
\cos^{2k}\theta\sin^{2(n-k)}\theta,
\end{equation}
for $\cos^2\theta=\bar{q}$. This means that
\begin{equation}
\frac{\partial f}{\partial x}(0,\ketbra{\Phi})=1-\sum_k\frac{p_k}{r_k}
|\Phi_k|^2 \ge 1-\sum_k\frac{p_k}{r_k} C^n_k
\cos^{2k}\theta\sin^{2(n-k)}\theta\ge 0,
\end{equation}
and that $\sigma^*$ is indeed the closest separable state to $\rho$ in
Eq.~(\ref{eqn:mixSnk}). Hence, Theorem 1 is proved.
\subsection{Examples}

We illustrate the  established expression of REE for the state $\rho(\{p\})$,
 making the restriction to mixtures of two distinct $n$-qubit states
$\ket{S(n,k_1)}$ and $\ket{S(n,k_2)}$ (with $k_1\ne k_2$):
\begin{eqnarray}
\rho_{n;k_1,k_2}(s)\equiv s\ketbra{S(n,k_1)} %\nonumber \\
+(1-s)\ketbra{S(n,k_2)}.
\end{eqnarray}
  We first investigate the two-qubit (i.e.\ $n=2$) case.
  Besides the trivial mixture, $\rho_{2;0,2}$, there is only one
inequivalent mixture, $\rho_{2;0,1}(s)$ [which is equivalent to
$\rho_{2;2,1}(s)$], which is the so-called {\it maximally entangled mixed
state\/}~\cite{MunroEtAl,WeiEtAl} (for a certain range of $s$)
\begin{equation}
\rho_{2;0,1}=s\,\ketbra{11}+(1-s)\ketbra{\Psi^+},
\end{equation}
where $\ket{\Psi^+}\equiv(\ket{01}+{10})/\sqrt{2}$ is one of the four Bell
states. The function $F$ for this state [denoted by $F_{2;0,1}(s)$] is
\begin{equation}
\label{eqn:rho201} F_{2;0,1}(s)=s
\,\log_2\frac{4s}{(1+s)^2}+(1-s)\log_2\frac{2}{1+s}\,,
\end{equation}
which is convex in $s$. Hence, it is exactly the expression for the relative
entropy of entanglement for the state $\rho_{2;0,1}$ found by Vedral and
Plenio~\cite{VedralPlenio98}.

For $n=3$ there are three nontrivial inequivalent mixtures: $\rho_{3;0,1}(s)$
[equivalent to $\rho_{3;3,2}(s)$], $\rho_{3;0,2}(s)$ [to $\rho_{3;3,1}(s)$],
and $\rho_{3;1,2}(s)$ [to $\rho_{3;2,1}(s)$]. In Fig.~\ref{fig:Er3} we compare
the function $F$ in Eq.~(\ref{eqn:F}), its convex hull  ${\rm co}\,F$, and
numerical values of $E_{\rm R}$ obtained using the general scheme described in
Ref.~\cite{VedralPlenio98} extended beyond the two-qubit case.

\begin{figure}%[t]
\centerline{\psfig{figure=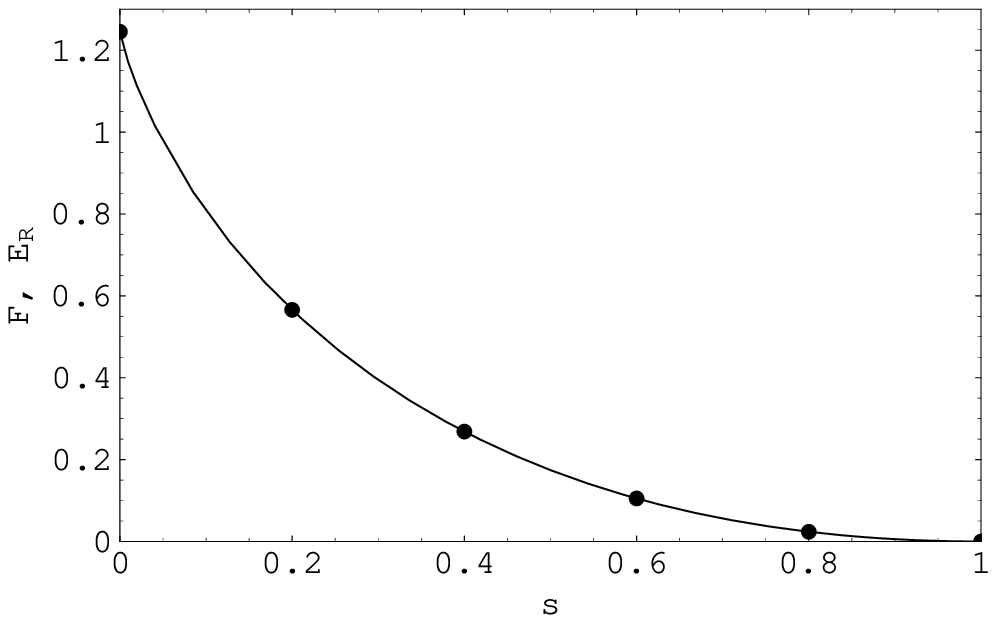,width=7cm,height=4cm,angle=0}}
\vspace{0.2cm}
\centerline{\psfig{figure=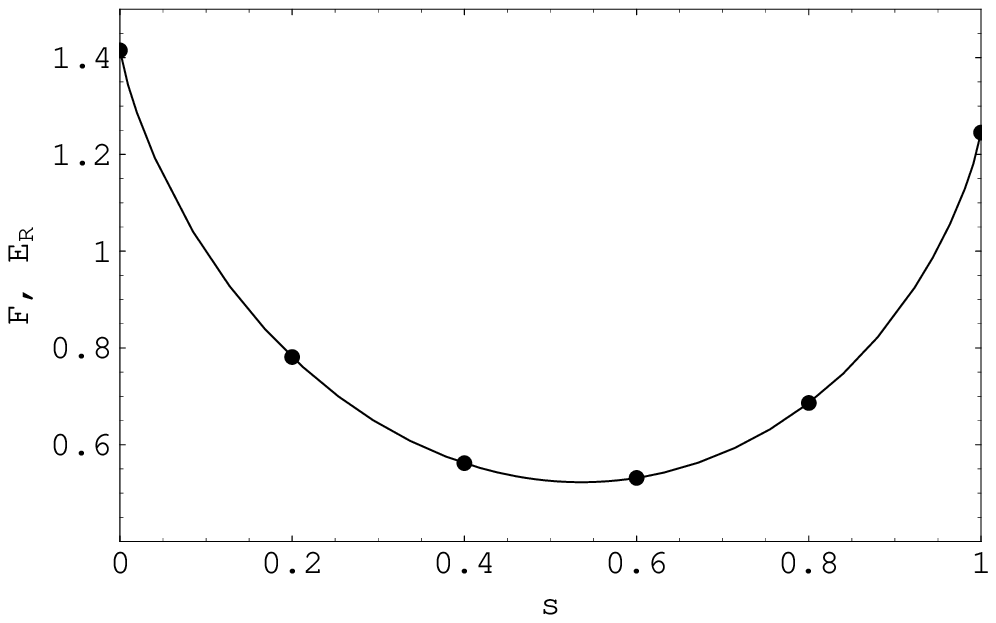,width=7cm,height=4cm,angle=0}}
\vspace{0.2cm}
\centerline{\psfig{figure=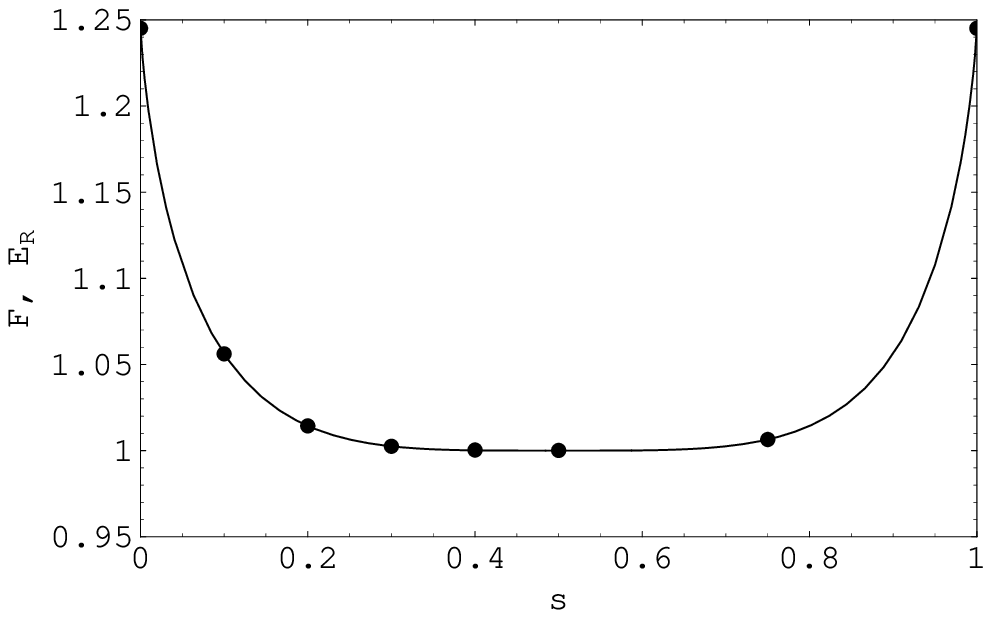,width=7.2cm,height=4cm,angle=0}}
%\vspace{- 0.2cm}
\caption{Comparison of $F$ (solid), its convex hull, i.e., $E_{\rm R}$  and
the numerical value of $E_{\rm R}$ for the states $\rho_{4;0,1}(s)$,
$\rho_{4;1,2}(s)$, and $\rho_{4;1,3}(s)$ (from top to bottom). In these cases,
$E_{\rm R}=F$.} \label{fig:Er4B}
%\vspace{-0.5cm}
\end{figure}

\begin{figure}%[t]
\centerline{\psfig{figure=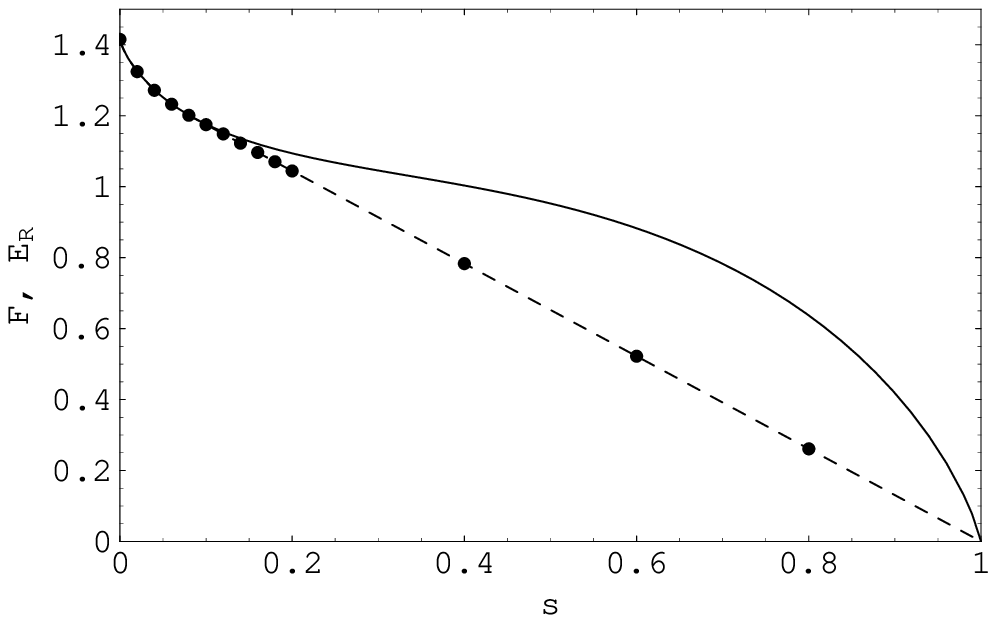,width=7cm,height=4cm,angle=0}}
\vspace{0.2cm}
\centerline{\psfig{figure=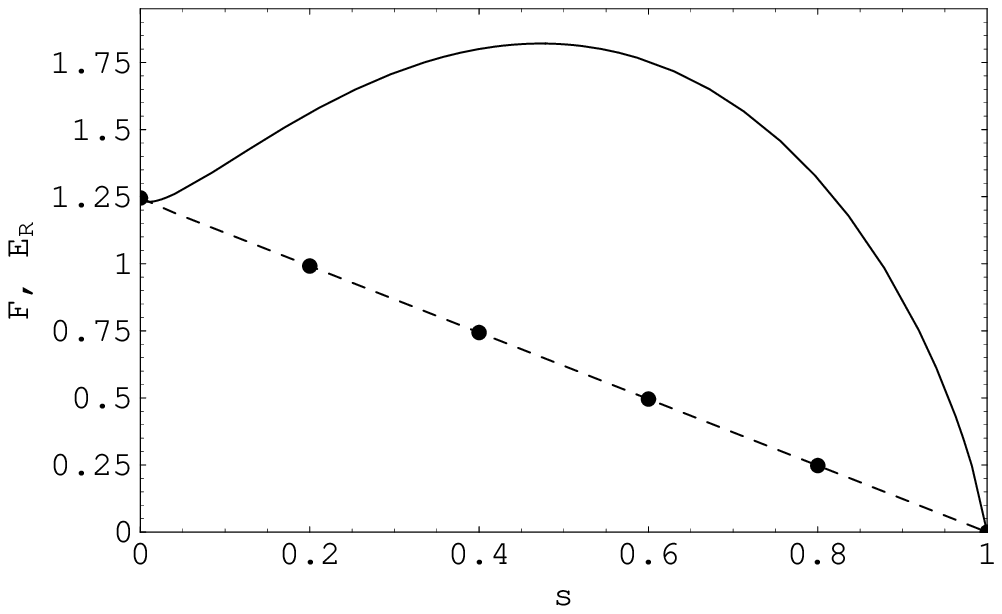,width=7cm,height=4cm,angle=0}}
%\vspace{- 0.2cm}
\caption{Comparison of $F$ (solid), its convex hull, i.e., $E_{\rm R}$ (partly
dashed line, when convexification is necessary), and the numerical value of
$E_{\rm R}$ for the states $\rho_{4;0,2}(s)$ (top) and
 $\rho_{4;0,3}(s)$ (bottom).
} \label{fig:Er4A}
%\vspace{-0.5cm}
\end{figure}

For $n=4$ there are five inequivalent nontrivial mixtures: $\rho_{4;0,1}(s)$,
$\rho_{4;0,2}(s)$, $\rho_{4;0,3}(s)$, $\rho_{4;1,2}(s)$, and
$\rho_{4;1,3}(s)$. In Figs.~\ref{fig:Er4A} and \ref{fig:Er4B} we again compare
the function $F$ in Eq.~(\ref{eqn:F}), its convex hull ${\rm co}\,F=E_{\rm
R}$, and numerical values of $E_{\rm R}$.

 For the states that we have just considered, we now explicitly give the
formulas for $E_{\rm R}$ suggested by the theorem.  For the three-qubit mixed
state $\rho_{3;2,1}(s)$, its $E_{\rm R}$ is
\begin{subequations}
\begin{equation}
\label{eqn:ErWW} s\log_2\frac{9s}{(1+s)^2(2-s)}+
(1-s)\log_2\frac{9(1-s)}{(2-s)^2(1+s)}.
\end{equation}
For $\rho_{3;0,1}(s)$, it is
\begin{equation}
\label{eqn:rho301} s\log_2\frac{27s}{(2+s)^3}+ (1-s)\log_2\frac{9}{(2+s)^2}.
\end{equation}
\end{subequations}
For $\rho_{4;0,1}(s)$, it is
\begin{subequations}
\begin{equation}
s\log_2\frac{256s}{(3+s)^4}+ (1-s)\log_2\frac{64}{(3+s)^3}.
\end{equation}
For $\rho_{4;1,2}(s)$, it is
\begin{equation}
s\log_2\frac{64s}{(2\!-\!s)(2\!+\!s)^3}+
(1\!-\!s)\log_2\frac{128(1-s)}{3(2\!-\!s)^2(2\!+\!s)^2}.
\end{equation}
For $\rho_{4;1,3}(s)$, it is
\begin{equation}
s\log_2\frac{64s}{(3\!-\!2s)(1\!+\!2s)^3}+
(1\!-\!s)\log_2\frac{64(1-s)}{(3\!-\!2s)^3(1\!+\!2s)}.
\end{equation}
\end{subequations}
These states above are exemplified in  Figs.~\ref{fig:Er3} and~\ref{fig:Er4B}.
For states such as $\rho_{3;0,2}$, $\rho_{4;0,2}$, and $\rho_{4;0,3}$,
convexifications (i.e. convex hull constructions) are needed; see
Figs.~\ref{fig:Er3} and~\ref{fig:Er4A}.

\subsection{Multi-qudits}
Now we consider multipartite qudit systems, in particular, the family of mixed
states
\begin{equation}
\label{eqn:Snkvecmixed} \rho(\{p_{\vec{k}}\})\equiv \sum_{\vec{k}\in {\cal
N}^d;\sum_i k_i=n}p_{\vec{k}}\ketbra{S(n;\vec{k})}.
\end{equation}
The closest separable state that is in the symmetric subspace is of the form
\begin{equation}
\label{eqn:csqudit} \sigma^*=\sum_{\vec{k}}r_{\vec{k}}\ketbra{S(n;\vec{k})},
\end{equation}
and it can be constructed from the convex hull of product states
\begin{equation}
\label{eqn:convex}
 \sum_i t_i \ketbra{\phi_i}^{\otimes n}.
\end{equation}
Let us now describe how it is constructed and how the formula for the REE is
obtained in a similar to the qubit case.
\begin{equation}
\ket{\Phi(\vec{u},\vec{\phi})}=\otimes_{j=1}^n \big(\sum_{l=1}^{d}
\sqrt{u_{l}}\,e^{i\phi_l}\ket{l}\big)_j=\sum_{\vec{k}}
\sqrt{\frac{n!}{k_1!\dots k_d!}}\big(\sqrt{u_1}\big)^{k_1}\dots
\big(\sqrt{u_d}\big)^{k_d}e^{i\phi_1+\dots+i\phi_d}\ket{S(n;\vec{k})}.
\end{equation}
Construct a separable mixed state by averaging over all phases,
\begin{equation}
\sigma(\vec{u})\equiv \int_0^{2\pi}d\phi_1\dots \int_0^{2\pi}d\phi_d
\ketbra{\Phi(\vec{u},\vec{\phi})}=\sum_{\vec{k}}C^n_{\vec{k}}\, u_1^{k_1}\dots
u_d^{k_d}\ketbra{S(n;\vec{k})},
\end{equation}
where, for the sake of convenience, we have defined $C^n_{\vec{k}}\equiv
n!/k_1!\dots k_d!$. The relative entropy between $\rho$ and $\sigma$ becomes
\begin{equation}
S(\rho\vert\vert\sigma)=\sum_{\vec{k}}
p_{\vec{k}}\log\frac{p_{\vec{k}}}{C^n_{\vec{k}}\, u_1^{k_1}\dots u_d^{k_d}}.
\end{equation}
Minimizing this w.r.t. $u$'s with the constraint $\sum_i u_i=1$, we arrive at
the solution
\begin{equation}
\bar{u}_j\equiv\frac{1}{n}\sum_{\vec{k}}p_{\vec{k}}k_j.
\end{equation}
This leads us to define the function
\begin{equation}
\label{eqn:Fqudit}
 F(p_{\vec{k}})\equiv\sum_{\vec{k}}
p_{\vec{k}}\log\frac{p_{\vec{k}}}{C^n_{\vec{k}}\, \bar{u}_1^{k_1}\dots
\bar{u}_d^{k_d}}.
\end{equation}
The relative entropy of entanglement for $\rho(p_{\vec{k}})$ is then
conjectured to be the convex hull of the above expression,\\
\noindent {\bf Theorem 2\/}:
\begin{equation}
E_{\rm R}\big(\rho(p_{\vec{k}})\big)={\rm co}\,F(p_{\vec{k}}),
\end{equation}
and the closest separable state will be the corresponding convex hull of
$\sigma(\vec{\bar{u}})$. Whenever the function $F$ is convex, there is no need
for the last convexification procedure, and $E_{\rm R}=F$.

Let us go on to prove this. As $\sigma^*$ in Eq.~(\ref{eqn:csqudit}) is
constructed as the closest separable state to $\rho(\{p_{\vec{k}}\})$ when
restricted in the symmetric subspace, we have in particular that
\begin{equation}
\frac{\partial f}{\partial x}(0,\ketbra{\Phi_s})\ge 0,
\end{equation}
for any $\ket{\Phi_s}$ of the form
\begin{equation}
\ket{\Phi_s}=\otimes_{j=1}^n \big(\sum_{l=1}^{d}
\sqrt{q_{l}}\ket{l}\big)_j=\sum_{\vec{k}}
\sqrt{C^n_{\vec{k}}}\big(\sqrt{q_1}\big)^{k_1}\dots
\big(\sqrt{q_d}\big)^{k_d}\ket{S(n;\vec{k})}.
\end{equation}
This means that
\begin{equation}
1-\sum_{\vec{k}}
\frac{p_{\vec{k}}}{r_{\vec{k}}}C^n_{\vec{k}}\big({q_1}\big)^{k_1}\dots
\big({q_d}\big)^{k_d}\ge0,
\end{equation}
for all $q_j\ge$ and $\sum_{j=1}^d q_j=1$. Similar to the qubit case, if we
can show that
\begin{equation}
\frac{\partial f}{\partial x}(0,\ketbra{\Phi})\ge 0,
\end{equation}
for arbitrary product state $\ket{\Phi}$,
\begin{equation}
\ket{\Phi}=\otimes_{j=1}^n \big(\sum_{l=1}^{d}
\sqrt{q_{j,l}}e^{i\phi_{j,l}}\ket{l}_j\big),
\end{equation}
for $\sum_{l=1}^d q_{j,l}=1$. We thus need to evaluate
\begin{equation}
\Phi_{\vec{k}}\equiv \ipr{S(n;\vec{k})}{\Phi}.
\end{equation}
Note that
\begin{equation}
\sqrt{ C^n_{\vec{k}}}\ket{S(n,\vec{k})}=\frac{1}{k_1!\dots k_d!}\sum_{\Pi_i
\in S_n}
\ket{\Pi_i(\underbrace{1,..,1}_{k_1},\underbrace{2,..,2}_{k_2},..,\underbrace{d,..,d}_{k_d})}.
\end{equation}
 This
gives
\begin{equation}
\Phi_{\vec{k}}=\frac{\sqrt{C^n_{\vec{k}}}}{n!}\sum_{\Pi_i}\sqrt{q_{\Pi_i(1),1}}e^{i\phi_{\Pi_i(1),1}}\dots\sqrt{q_{\Pi_i(k_1),1}}e^{i\phi_{\Pi_i(k_1),1}}
\sqrt{q_{\Pi_i(k_1+1),2}}e^{i\phi_{\Pi_i(k_1+1),2}}\dots\sqrt{q_{\Pi_i(n),d}}e^{i\phi_{\Pi_i(n),d}}.
\end{equation}
Taking the absolute value, we have
\begin{equation}
|\Phi_{\vec{k}}|\le\frac{\sqrt{C^n_{\vec{k}}}}{n!}\sum_{\Pi_i}\big(\sqrt{q_{\Pi_i(1),1}}\dots\sqrt{q_{\Pi_i(k_1),1}}\big)
\big(\sqrt{q_{\Pi_i(k_1+1),2}}\dots\sqrt{q_{\Pi_i(k_1+k_2),2}})\dots
\big(\sqrt{q_{\Pi_i(n-k_d+1),d}}\dots\sqrt{q_{\Pi_i(n),d}}\big).
\end{equation}

Recently, Carlen, Loss and Lieb~\cite{CarlenLossLieb06} have shown an
inequality regarding the permanent of a matrix. Using their results, the
following inequality is easily seen to hold:
\begin{equation}
\frac{1}{n!}\sum_{\Pi_i}\big(\sqrt{q_{\Pi_i(1),1}}\dots\sqrt{q_{\Pi_i(k_1),1}}\big)
\big(\sqrt{q_{\Pi_i(k_1+1),2}}\dots\sqrt{q_{\Pi_i(k_1+k_2),2}})\dots
\big(\sqrt{q_{\Pi_i(n-k_d+1),d}}\dots\sqrt{q_{\Pi_i(n),d}}\big)\le\big({\overline{q_1}}\big)^{k_1/2}\dots
\big(\overline{q_d}\big)^{k_d/2},
\end{equation}
where $\overline{q_l}=\sum_{j=1}^n q_{j,l}/n$. With this, our conjecture is
thus proved.

\subsection{Examples}
\begin{figure}%[t]
\centerline{\psfig{figure=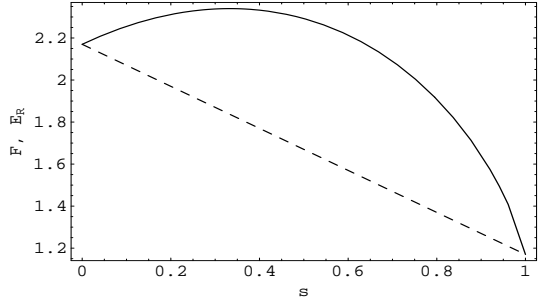,width=7.2cm,height=4cm,angle=0}}
%\vspace{- 0.2cm}
\caption{Comparison of $F$ (solid), its convex hull, i.e., $E_{\rm R}$
(dashed) for the state $\rho_{ab}(s)$. } \label{fig:Erqudit}
%\vspace{-0.5cm}
\end{figure}
Let us first at a three-qutrit example: $n=3$, $d=4$, and involving only two
vectors $\vec{a}=(2,0,0,1)$ and $\vec{b}=(1,1,1,0)$, to which  the
corresponding states are
\begin{eqnarray}
\label{eqn:a}
\ket{\vec{a}}&\equiv&\frac{1}{3}\big(\ket{114}+\ket{141}+\ket{411}\big),\\
\label{eqn:b}
\ket{\vec{b}}&\equiv&\frac{1}{6}\big(\ket{123}+\ket{132}+\ket{213}+\ket{231}+\ket{312}+\ket{321}\big).
\end{eqnarray}
From Eq.~(\ref{eqn:equalityqudit}) their REE are $2\log_2(3)-2\approx 1.17$
and $2\log_2(3)-1\approx2.17$, respectively. We shall now consider the mixture
\begin{equation}
\rho_{ab}(s)\equiv s\ketbra{\vec{a}}+(1-s)\ketbra{\vec{b}}.
\end{equation}
According to Eq.~(\ref{eqn:Fqudit}), the corresponding $F$ function is
\begin{equation}
F_{ab}(s)=s \log_2\frac{9}{(1+s)^2}+(1-s)\log_2\frac{9}{2(1-s^2)},
\end{equation}
which is, however, not convex. This means that we need to construct its convex
hull in order to obtain its REE,
\begin{equation}
E_{\rm R}(s)={\rm co}\, F_{ab}(s)=F_{ab}(1) + (1 - s)(F_{ab}(0) - F_{ab}(1)).
\end{equation}
This is shown in Fig.~\ref{fig:Erqudit}.

As another example, let us consider additionaly the state assocaited with
${\vec{c}}=(1,0,0,2)$ defined via
\begin{equation}
\ket{\vec{c}}\equiv\frac{1}{3}\big(\ket{441}+\ket{414}+\ket{144}\big).\\
\end{equation}
The state mixture of $\ket{\vec{a}}$ and $\ket{\vec{c}}$,
\begin{equation}
\rho_{ac}(s)=s\ketbra{\vec{a}}+(1-s)\ketbra{\vec{c}},
\end{equation}
can be seen to possess REE equal to that of $\rho_{3;1,2}(s)$.
\section{$E_{\log}\le E_{\rm R}$?}

 Recall that for pure states we found the
inequality $E_{\log}\le E_{\rm R}$. Does this inequality hold for mixed
states? We do not know the complete answer to this question. From Inequality 6, we only have
$E_{\log}(\rho)-S(\rho)\le E_{\rm R}(\rho)$.  However, we shall show that
$E_{\log}\le E_{\rm R}$ indeed holds for the two families of mixed states in Eqs.~(\ref{eqn:mixSnk}) and~(\ref{eqn:Snkvecmixed}), as well as Eq.~(\ref{eqn:DurW}).

\subsection{Multi-qubits}
Let us begin with the multi-qubit states~(\ref{eqn:mixSnk}). We first establish that the quantity ${\cal E}(\{q\})$ that is used to obtain $E_{\log}$ is
a lower bound on $F(\{q\})$, which is used to obtain $E_{\rm R}$.
Recall that~\cite{WeiEricssonGoldbartMunro04}
\begin{subequations}
\begin{equation}
{\cal E}(\{p\})=
-2\log_2\left[\max_\theta \sum_k \sqrt{p_k}\, \sqrt{C_k^n} \cos^k\theta\sin^{n-k}\theta\right].
\end{equation}
By the concavity of $\log$, we then have
\begin{eqnarray}
-2\log_2\left[\sum_k \sqrt{p_k}\, \sqrt{C_k^n} \cos^k\theta\sin^{n-k}\theta\right] %\nonumber \\
\le \sum_k p_k \log_2\frac{p_k}{C_k^n \cos^{2k}\theta \sin^{2(n-k)}\theta}.
\end{eqnarray}
Hence
\begin{eqnarray}
\min_\theta -2\log_2\left[\sum_k \sqrt{p_k}\, \sqrt{C_k^n} \cos^k\theta\sin^{n-k}\theta\right] %\nonumber \\
\le \min_\theta\sum_k p_k \log_2\frac{p_k}{C_k^n \cos^{2k}\theta \sin^{2(n-k)}\theta},
\end{eqnarray}
or equivalently
\begin{equation}
{\cal E}(\{p\})\le F(\{p\}).
\end{equation}
\end{subequations}
 By using theorem 1 and by taking the convex hull of both sides
of this inequality we  have
\begin{equation}
E_{\log}\le E_{\rm R}
\end{equation}
for the family of states~(\ref{eqn:mixSnk}).

\subsection{Multi-qudits}
Similarly, the relation $E_{\log}\le E_{\rm R}$ also holds
true for the multi-qudit states~(\ref{eqn:Snkvecmixed}), and the proof is similar.

Following the same idea in Ref.~\cite{WeiEricssonGoldbartMunro04}, to calculate $E_{\log}$ for states~(\ref{eqn:Snkvecmixed}), one first consider
\begin{subequations}
\begin{equation}
{\cal E}(\{p_{\vec{k}}\})=
-2\log_2\left[\max_{\vec{u}} \sum_{\vec{k}} \sqrt{p_{\vec{k}}}\, \sqrt{C_{\vec{k}}^n} (\sqrt{u_1})^{k_1}\dots(\sqrt{u_d})^{k_d}\right].
\end{equation}
By the concavity of $\log$, we then have
\begin{eqnarray}
-2\log_2\left[\sum_{\vec{k}} \sqrt{p_{\vec{k}}}\, \sqrt{C_{\vec{k}}^n} (\sqrt{u_1})^{k_1}\dots(\sqrt{u_d})^{k_d}\right] %\nonumber \\
\le \sum_{\vec{k}} p_{\vec{k}} \log_2\frac{p_{\vec{k}}}{C_{\vec{k}}^n {u_1}^{k_1}\dots{u_d}^{k_d}}.
\end{eqnarray}
Hence
\begin{eqnarray}
\min_{\vec{u}} -2\log_2\left[\sum_{\vec{k}} \sqrt{p_{\vec{k}}}\, \sqrt{C_{\vec{k}}^n} (\sqrt{u_1})^{k_1}\dots(\sqrt{u_d})^{k_d}\right] %\nonumber \\
\le \min_{\vec{u}}\sum_{\vec{k}} p_{\vec{k}} \log_2\frac{p_{\vec{k}}}{C_{\vec{k}}^n {u_1}^{k_1}\dots{u_d}^{k_d}},
\end{eqnarray}
or equivalently
\begin{equation}
{\cal E}(\{p_{\vec{k}}\})\le F(\{p_{\vec{k}}\}).
\end{equation}
\end{subequations}
 By using theorem 1 and by taking the convex hull of both sides
of this inequality we have
\begin{equation}
E_{\log}\le E_{\rm R}
\end{equation}
for the family of states~(\ref{eqn:mixSnk}).

It would be interesting to know to what extent $E_{\log}\le E_{\rm R}$ holds.

\section{Some applications }
\label{sec:applications}
 We know that the W state $\ket{W}\equiv(\ket{001}+\ket{010}+\ket{100})/\sqrt{3}$
 is more robust than $\ket{GHZ}\equiv(\ket{000}+\ket{111})/\sqrt{2}$ against
 losing one of their constituent parties. This is also true for
 the particular family of $n$-qubit pure states $\{\ket{S(n,k)}\}$,
the relative entropy of entanglement of which we have given in Eq.~(\ref{eqn:rhoSnk}).
Now, if we trace over one party we get a mixed $(n-1)$-qubit state:
\begin{equation}
\label{eqn:Tr1}
{\rm Tr}_1 \ketbra{S(n,k)}=\frac{n\!-\!k}{n}\ketbra{S(n\!-\!1,k)}+\frac{k}{n}\ketbra{S(n\!-\!1,k\!-\!1)}.
\end{equation}
We have also given an expression for the relative entropy of entanglement for
this mixed state.  If we trace over $m$ parties, the reduced mixed state would
be a mixture of $\{\ket{S(n-m,q)}\}$ [with $q\le (n-m)$], and again we have
its relative entropy of entanglement.  For example, if we start with
$\ket{S(4,1)}$, and trace over one party and then another, we get the
sequence:
\begin{equation}
\ket{S(4,1)}\rightarrow \rho_{3;0,1}({1}/{4})
\rightarrow
\rho_{2;0,1}({1}/{2}),
\end{equation}
for which we have given the corresponding relative entropies of entanglement
in Eqs.~(\ref{eqn:rhoSnk}), (\ref{eqn:rho301}) and (\ref{eqn:rho201}). The
entanglement at each stage is
\begin{equation}
3\log_2\left(\frac{4}{3}\right)\rightarrow \log_2\left(\frac{16}{9\cdot
3^{1/4}}\right) \rightarrow \frac{1}{2}\log_2\left(\frac{32}{27}\right),
\end{equation}
or numerically,
\begin{equation}
1.24511\rightarrow 0.433834 \rightarrow 0.122556.
\end{equation}
Unlike $n$-GHZ states, these states can still remain entangled even if some of
the qubits are lost along the way.

Plenio and Vedral~\cite{PlenioVedral01} have derived a lower bound on the REE
of a tripartite pure state $\rho_{\rm ABC}=\ketbra{\psi}$ in terms of the the
entropies and REE of the reduced states of two parties:
\begin{equation}
\max\{E_{\rm R}(\rho_{\rm AB})+S(\rho_{\rm AB}),E_{\rm R}(\rho_{\rm AC})+S(\rho_{\rm AC}),E_{\rm R}(\rho_{\rm BC})+S(\rho_{\rm BC})\}\le E_{\rm R}(\rho_{\rm ABC}),
\end{equation}
where $\rho_{\rm AB}={\rm Tr}_C(\rho_{\rm ABC})$ (and similarly for $\rho_{\rm
AC}$ and $\rho_{\rm BC}$) and $S(\rho)\equiv - {\rm Tr}\rho\log_2\rho$ is the
von Neumann entropy. They have further found that this lower bound is
saturated by $\ket{\rm GHZ}$ and $\ket{\rm W}$.  Based on  Theorem~1, we can
actually show that for $\rho_{12\ldots n}=\ketbra{S(n,k)}$ the inequality
\begin{equation}
\label{eqn:CRE}
\max_i \{E_{\rm R}(\rho_{12\ldots\hat{i}\ldots n})+S(\rho_{ 12\ldots\hat{i}\ldots n})\}\le E_{\rm R}(\rho_{12\ldots n})
\end{equation}
is saturated, where $\rho_{12\ldots\hat{i}\ldots n}\equiv{\rm
Tr}_i(\rho_{12\ldots n})$ is the reduced density matrix obtained from
$\rho_{12\ldots n}$ by tracing out the $i$-th party. The proof is as follows.
As $\ket{S(n,k)}$ is permutation-invariant, there is no need to maximize over
all parties, and we can simply take $i=1$, obtaining the reduced state
$\rho_{n-1;k-1,k}(k/n)$ as in Eq.~(\ref{eqn:Tr1}). As the corresponding
function $F_{n-1;k-1,k}(s)$ of $\rho_{n-1;k-1,k}(s)$ is convex for
$s\in[0,1]$, we immediately obtain from Theorem~1 that, for
$\rho_{n-1;k-1,k}(k/n)$,
\begin{subequations}
\begin{eqnarray}
\!\!\!\!\!\!\!\!\!\!\!E_{\rm
R}\left(\rho_{n\!-\!1;k\!-\!1,k}(k/n)\right)&=&-\log_2\left[C^n_k
\left(\frac{k}{n}\right)^k\left(\frac{n\!-\!k}{n}\right)^{n\!-\!k}\right]
+\frac{k}{n}\log_2\frac{k}{n}+\frac{n\!-\!k}{n}\log_2\frac{n\!-\!k}{n}\\
&=&E_{\rm R}\left(\ket{S(n,k)}\right)- S\left(\rho_{n\!-\!1;k\!-\!1,k}(k/n)\right).
\end{eqnarray}
\end{subequations}
Therefore, the bound in Eq.~(\ref{eqn:CRE}) is saturated for $\rho_{12\ldots
n}=\ketbra{S(n,k)}$.

We remark that there are other applications that our results can be useful,
such as (i) providing bounds on state discrimination by separable operations
and (ii) constructing optimal entanglement witness, which have been discussed
by Hayashi et al.~\cite{HayashiMarkhamMuraoOwariVirmani08}, and we refer
readers to their paper.
\section{Another family of multi-qubit mixed states: D\"ur's states}
\label{sec:Dur}
 D\"ur~\cite{Dur01} found that for $N\ge 4$ the
following state is bound entangled:
\begin{equation}
\rho_N\equiv\frac{1}{N+1}\left(\ketbra{\Psi_G}+\frac{1}{2}\sum_{k=1}^N\big(P_k+\bar{P}_k\big)\right),
\end{equation}
where $\ket{\Psi_G}\equiv \big(\ket{0^{\otimes
N}}+e^{i\alpha_N}\ket{1^{\otimes N}}\big)/        {\sqrt{2}}$ is a $N$-partite
GHZ state; $P_k\equiv\ketbra{u_k}$ is a projector onto the state
$\ket{u_k}\equiv\ket{0}_1\ket{0}_2\ldots\ket{1}_k\ldots\ket{0}_N$; and
$\bar{P}_k\equiv\ketbra{v_k}$ projects on to
$\ket{v_k}\equiv\ket{1}_1\ket{1}_2\ldots\ket{0}_k\ldots\ket{1}_N$. For $N\ge
8$ this state violates the (two-setting) Mermin-Klyshko-Bell
inequality~\cite{Dur01}; violation was pushed down to $N\ge 7$ by Kaszlikowski
et al.~\cite{Kwek02} for a three-setting Bell inequality; it was pushed
further down to $N\ge 6$ by Sen {\it et al.\/}~\cite{SenSenZukowski02} for a
functional Bell inequality. The phase $\alpha_N$ in $\ket{\Psi_G}$ can be
eliminated by local unitary transformations, and hence we shall take
$\alpha_N=0$ in the following discussion.

This state has been extended to the following family by Wei et
al.~\cite{WeiAltepeterGoldbartMunro04},
\begin{equation}
\label{eqn:DurW}
 \rho_N(x)\equiv x\ketbra{\Psi_G}+\frac{1-x}{2N}
\sum_{k=1}^N\big(P_k+\bar{P}_k\big),
\end{equation}
and they found that for $N\ge 4$ the state is bound entangled if $0<x\le
1/(N+1)$ and is still entangled but not bound entangled if $ 1/(N+1)<x\le1$.
They calculated the negativity and the geometric measure for this family of
states, e.g.,
\begin{equation}
E_{\log}\big(\rho_N(x)\big)=\log_2\frac{2}{2-x},
\end{equation}
and conjectured
their relative entropy of entanglement to be $E_R(x)=x$, for $N\ge 4$, with
one closest separable mixed state being
%\begin{widetext}
\begin{equation}
\sigma^*(x)=\frac{x}{2}\big(\ketbra{\Psi_G}+\ketbra{\Psi_G^-}\big)
+\frac{1-x}{2N}\sum_{k=1}^N\big(P_k +\bar{P}_k\big),
\end{equation}
%\end{widetext}
where
\begin{equation}
\ket{\Psi_G^-}\equiv\frac{1}{\sqrt{2}} \big(\ket{0^{\otimes
N}}-\ket{1^{\otimes N}}\big).
\end{equation}
 We prove their conjecture here.
 As before we define the quantity
 \begin{equation}
f(z,\sigma_s)\equiv S(\rho_N(x)|| (1-z)\sigma^*(x)+ z \sigma_s),
\end{equation}
where $\sigma_s$ is any separable state. In order to show that $\rho^*(x)$ is
indeed the closest separable state to $\rho_N(x)$, it is sufficient to show
that
\begin{equation}
\frac{\partial f(0,\ketbra{\Phi})}{\partial z}=1-\int_0^\infty dt\, {\rm
Tr}\big[ (\sigma^*+t)^{-1}\rho_N (\sigma^*+t)^{-1}\ketbra{\Phi}\big]\ge 0,
\end{equation}
where $\ket{\Phi}$ is any separable pure state, which can be parametrized as
follows,
\begin{equation}
\ket{\Phi}=\otimes_{j=1}^n(\cos\theta_j\ket{0}_j+ \sin\theta_j\,
e^{i\phi_j}\ket{1}_j),
\end{equation}
and, without loss of generality, we can restrict ourselves to
$\cos\theta_j,\,\sin\theta_j\ge0$. By direct calculation, we have
\begin{equation}
\int_0^\infty dt\,(\sigma^*+t)^{-1}\rho_N (\sigma^*+t)^{-1}={2}\ketbra{\Psi_G}
+\sum_{k=1}^N\big(P_k +\bar{P}_k\big),
\end{equation}
and hence
\begin{equation}
\frac{\partial f(0,\ketbra{\Phi})}{\partial z}=1-g_N\ge 0
\end{equation}
is valid for $N\ge 4$ , as (see Ref.~\cite{WeiAltepeterGoldbartMunro04} for
the proof)
\begin{eqnarray}
&& g_N\equiv\big(c_1\cdots c_N+s_1\cdots s_N\big)^2
+\nonumber \\
&&\quad\sum_{k=1}^N\left\{ (c_1\cdots s_k\cdots c_N)^2 +(s_1\cdots c_k\cdots
s_N)^2\right\}\le 1,
\end{eqnarray}
where we have simplified the notation by using $c_i\equiv \cos\theta_i$ and
$s_i\equiv \sin\theta_i$. Therefore, $\sigma^*(x)$ is indeed the closest
separable state to $\rho_N(x)$ and hence $E_R\big(\rho_N(x)\big)=x$ for $N\ge
4$.

We remark that for this family of states, the relation $E_{\log}\le E_{\rm R}$
holds, as $\log_2\big(2/(2-x)\big)\le x$ for $0\le x\le 1$.
\section{Concluding remarks}
\label{sec:summary} We  have proved conjectures on the relative entropy of
entanglement (REE) for two families of multipartite  qubit states. Thus,
analytic expressions of REE for these families of states can be
straightforwardly obtained. The first family of states we consider are
composed of mixture of some permutation-invariant multi-qubit states. We have
also generalized the results generalized to permutation-invariant multi-qudit
states, and have given the expression for their relative entropy of
entanglement.  The second family of states contain D\"ur's multipartite bound
entangled states. Along the way, we have reviewed inequalities connecting the
relative entropy of entanglement to the robustness of entanglement and the
geometric measure of entanglement, and have slightly extended previous
discussions.

 It is possible that our results on the relative entropy of
entanglement can applied to the checking of the consistency of some equalities
and inequalities~\cite{PlenioVedral01,WuZhang00,GalvaoPlenioVirmani00}
regarding minimal reversible entanglement generating sets (MREGSs).  These
equalities and inequalities concerning MREGS usually involve only the von
Neumann entropy and the regularized (i.e.~asymptotic) relative entropy of
entanglement of the pure state and its reduced density matrices. The results
we have in the present Paper concern only the non-regularized version of the
relative entropy of entanglement, and hence, can only reach weaker conclusion.
Therefore, a major challenge is to extend the ideas contained in the present
Paper to the considerations of the regularized version of the relative entropy
of entanglement. Moreover, knowledge of the regularized version is also
important for providing bounds on the random bipartite entanglement recently
investigated by Fortescue and Lo~\cite{FortescueLo07}.

In Ref.~\cite{WeiGoldbart03} the geometric measure of entanglement was known
for not only the family of mixed states $\sum_k p_k\ketbra{S(n,k)}$ but also
the family of pure states (formed by superposition)
$\sum_k\alpha_k\ket{S(n,k)}$. For the relative entropy of entanglement, we
still do not yet have analytic expressions for the latter, except for the
basis states $\ket{S(n,k)}$. It will be desirable to consider how we can
obtain the relative entropy of entanglement for this family of pure
multi-qubit states, as well as the corresponding family of pure multi-qudit
states $\sum_{\vec{k}}\alpha_{\vec{k}}\ket{S(n,\vec{k})}$.

Indeed, investigating the relative entropy of entanglement for states in this
family $\sum_{\vec{k}}\alpha_{\vec{k}}\ket{S(n,\vec{k})}$ can serve as a first
step towards obtaining the regularized version of the relative entropy of
entanglement for states such as $\ket{S(n,k)}$ and $\ket{S(n,\vec{k})}$, as we
now illustrate. Let us consider two copies of the $W$ state (shared among
parties A, B, and C),
\begin{equation}
\ket{W^{\otimes 2}}=\ket{W}_{A_1B_1C_1}\otimes\ket{W}_{A_2B_2C_2},
\end{equation}
where
\begin{equation}
\ket{W}=\frac{1}{\sqrt{3}}(\ket{001}+\ket{010}+\ket{100}).
\end{equation}
We can expand $\ket{W^{\otimes 2}}$ and re-label the states as follows
\begin{subequations}
\begin{eqnarray} \ket{0}_{A_1}\otimes\ket{0}_{A_2}&\rightarrow& \ket{1}_A,\\
\ket{0}_{A_1}\otimes\ket{1}_{A_2}&\rightarrow& \ket{2}_A,\\
\ket{1}_{A_1}\otimes\ket{0}_{A_2}&\rightarrow& \ket{3}_A,\\
\ket{1}_{A_1}\otimes\ket{1}_{A_2}&\rightarrow& \ket{4}_A,
\end{eqnarray}
\end{subequations}
and similarly for parties B and C. This gives us
\begin{equation}
\ket{W^{\otimes
2}}=\frac{1}{\sqrt{3}}\ket{\vec{a}}_{ABC}+\frac{\sqrt{2}}{\sqrt{3}}\ket{\vec{b}}_{ABC},
\end{equation}
where $\ket{\vec{a}}$ and $\ket{\vec{b}}$ are defined in Eqs.~(\ref{eqn:a})
and~(\ref{eqn:b}), respectively. Thus, the two copies of the W-state is
identical to a superposition of two four-level three-party states, exactly of
the form $\sum_{\vec{k}}\alpha_{\vec{k}}\ket{S(n,\vec{k})}$. In general, a
state with $m$ copies of, say, $\ket{S(n,k)}$ is also of this form, and hence,
the knowledge of the relative entropy of entanglement for the former enables
the knowledge of the regularized relative entropy of entanglement for the
latter.

 In discussing the
symmetry of the separable states, we have essentially characterized the family
of states invariant under such symmetry (${\rm P}_1$ and ${\rm P}_2$). It
would be desirable to quantify the entanglement of such family. This family
includes states that are generalizations of Werner states, which invariant
under average action of $U(d)^{\otimes n}$ in multipartite systems. An
interesting question arises: can we characterize the distillability of
multipartite states just as in bipartite case, where the question can be
reduced to that for Werner states?

\section*{Acknowledgments}
\noindent We thank Tobias Moroder, Robert Raussendorf, Simone Severini and
Rolando Somma for useful discussions. This work was supported by IQC, NSERC,
and ORF.

\end{document}